\documentclass[hyper]{JHEP3}

\input{epsf}
\usepackage{epsfig}
\usepackage{amssymb}
\usepackage{amsfonts}
\usepackage{amsbsy}
\usepackage[all,2cell]{xy}

\usepackage{amsmath}

\usepackage{amssymb,amscd}
\usepackage{mathrsfs}
\usepackage{comment}
\usepackage{amsmath,amsthm}

\def\be{\begin{equation}}
\def\ee{\end{equation}}

\def\mod{{\rm mod}}
\def\Tr{{\rm Tr}}

\def\IC{\mathbb{C}}

\def\IP{\mathbb{P}}

\def\IR{{\mathbb{R}}}

\def\IZ{{\mathbb{Z}}}

\def\CC {{\cal C}}

\def\CN {{\cal N}}
\def\CR {{\cal R}}

\def\CF {{\cal F}}

\def\CV {{\cal V}}

\def\CO {{\cal O}}

\def\CE {{\cal E}}

\def\CH {{\cal H}}

\def\CQ {{\cal Q}}

\def\half{\frac{1}{2}}

\renewcommand{\Im}{{\rm Im }}

\def\one{{\hbox{ 1\kern-.8mm l}}}
\def\ch{{\rm ch \, }}

\def\be{\bar{e}}

\def\ch{{\rm ch}}

\def\half{\frac{1}{2}}

\def\ch{{\rm ch}}

\def\half{\frac{1}{2}}

\def\be{ \begin{equation} }
\def\ee{ \end{equation}}

\def\fB{\mathfrak{B}}

\def\I{{\rm i}}

\title{The Partition Function Of Argyres-Douglas Theory On A Four-Manifold}

\author{ Gregory W.~Moore and Iurii Nidaiev \\
NHETC and
$~~$Department of Physics and Astronomy, Rutgers University \\
$~~$126 Frelinghuysen Rd., Piscataway NJ 08855, USA\\
\\
{\tt gmoore@physics.rutgers.edu, iurii.nidaiev@gmail.com } }

\abstract{Using the $u$-plane integral as a tool, we derive a formula for the partition function of the simplest nontrivial (topologically twisted) Argyres-Douglas theory on compact, oriented,
simply connected, four-manifolds without boundary and with $b_2^+>0$.
The result can be expressed in terms of classical cohomological invariants
and Seiberg-Witten invariants. Our results hint at the existence of standard
four-manifolds that are not of Seiberg-Witten simple type.
\newline
\newline
 \today }

\begin{document}

\section{Introduction And Conclusion}

One of the great moments in the history of Physical Mathematics is
Witten's formulation  \cite{Witten:1994cg} of the Seiberg-Witten invariants of four-manifolds
together with his proposal for how to express the Donaldson invariants of four-manifolds
in terms of the Seiberg-Witten invariants - a result known to four-manifold
theorists as ``the Witten conjecture.''  The introduction of Seiberg-Witten invariants
led to rapid progress in the theory of four-manifolds. See Donaldson's review for a
masterful account \cite{DonaldsonReview}.
Nevertheless,   many interesting  questions in the field remain open \cite{FintushelStern,Scorpan,Stern}.

Given Witten's remarkable application of the basic Seiberg-Witten solution of pure
$SU(2)$ $N=2$ supersymmetric Yang-Mills theory \cite{Seiberg:1994rs} to four-manifold theory,  one naturally asks
whether topological twisting of other supersymmetric
quantum field theories will lead to other new four-manifold invariants.
One evident hunting ground in the search for such new invariants is the set of topologically twisted
four-dimensional $N=2$ theories. This was, in fact, the main motivation
for works such as \cite{Marino:1998bm}. While far from definitive, the
main conclusion of \cite{Marino:1998bm} was that, for Lagrangian theories,
the partition functions of the topologically twisted theories, while
intricate and interesting, will nevertheless be expressible in terms
of the classical topological invariants and Seiberg-Witten invariants of a
four-manifold. This narrows the search for new invariants to non-Lagrangian superconformal
theories. Again, this was the motivation for \cite{Marino:1998tb,Marino:1998uy}.
Those papers again failed to discover new invariants, but they did manage to show
 that the very existence of superconformal theories is related to  nontrivial sum rules on the
Seiberg-Witten invariants, now known as the ``superconformal simple type condition.''

The superconformal theory used in \cite{Marino:1998tb,Marino:1998uy} is the
simplest nontrivial Argyres-Douglas theory and is denoted here as $AD3$ (it is sometimes also
 denoted as the $(A_1,A_2)$ theory). It arises in the Coulomb branch of
 pure $SU(3)$ SYM \cite{Argyres:1995jj} and in the Coulomb branch of
 $SU(2)$ SYM coupled to a single hypermultiplet in the
fundamental representation \cite{Argyres:1995xn}.  The present
paper completes the story of \cite{Marino:1998tb,Marino:1998uy}
by giving an explicit formula for the partition function of the
topologically twisted AD3 theory on compact, oriented,
four-manifolds without boundary,  henceforth denoted by $X$, in the case that
$b_1(X)=0$ and $b_2^+(X)>0$. Four-manifolds of this type that satisfy the
further condition that
$b_2^+(X)>1$ are typically referred to  as \emph{standard four-manifolds}.
We will argue that the partition function of the twisted $AD3$ theory
on  standard four-manifolds
can, once again, be expressed using the classical topological invariants
and the Seiberg-Witten invariants. Our proposal for this partition function
 on standard four-manifolds is given by equation \eqref{eq:AD3-StandardX}.
While this partition function does not provide new four-manifold
invariants it might nevertheless be useful. For example, when  $X$ is of Seiberg-Witten simple type
\footnote{ A standard four-manifold
$X$ is said to be of \emph{Seiberg-Witten simple type} if the Seiberg-Witten
invariant associated to a spin-c structure is only nonvanishing when the moduli space of solutions to the
Seiberg-Witten equations is of dimension zero. For mathematical discussions
see \cite{DonaldsonReview,Morgan}.}
the formula simplifies dramatically to equation \eqref{eq:SimpleAD3}. (See Section
\ref{subsec:Discussion} for a discussion of this simplification.)   This brings into   sharp focus
the distinction between those manifolds of Seiberg-Witten simple type and those hypothetical
manifolds that are not of Seiberg-Witten simple type. In particular, equation \eqref{eq:SimpleAD3}
has the remarkable property that the $0$-observable is a ``null-vector'' in the sense that insertions
of this operator always lead to zero correlation function. That property is not true of the more
general expression \eqref{eq:AD3-StandardX}. Strangely enough, all
known standard four-manifolds are of Seiberg-Witten simple type. Why this should be so
is mysterious, at present. There ought to be a
good physical reason why the zero observable is a null vector. In the absence of such a
reason one must suspect that, actually, there do exist standard four-manifolds that are
not of Seiberg-Witten simple type.

As explained in Section \ref{sec:Continuous}, on manifolds with $b_2^+(X)=1$ the topologically
twisted $AD3$ partition function is, in fact,
not a diffeomorphism invariant, but varies \underline{continuously} with the metric. This is probably
a general feature of twisted superconformal field theories. The basic reason that lies behind this
failure of the general expectations of topological field theory was first noted for the
$SU(2)$, $N_f=4$ theory in \cite{Moore:1997pc}.

Our result does not imply that other
topologically twisted N=2 theories won't lead to new four-manifold invariants,
although it might dim the ardor of those in pursuit of such new invariants.
Whether or not other theories lead to new invariants, the computation of these partition functions
remains an interesting challenge for the future. Among other things, it would be of
interest to apply our methods to the other theories with one-dimensional Coulomb branches
described in \cite{Argyres:2015ffa,Argyres:2015gha,Argyres:2016xmc}.

Our basic line of reasoning is the following. From reference \cite{Argyres:1995xn}
we know that in the $SU(2)$ $N_f=1$ theory at a special value of the quark mass, $m=m_*$,
the IR physics near a special vacuum $u=u_*$ is described by the AD3 theory.
Moreover there is no noncompact Higgs branch for the $SU(2)$ $N_f=1$ theory
so if we take the $m \to m_*$ limit of the partition function we should be able
to extract the AD3 partition function. In section \ref{sec:Deriving}  below we will make this
reasoning a little more precise, and Appendix \ref{app:Zutilde} carries out the
procedure in great detail.  Our derivation makes use of a very general relation of $u$-plane
integrands to total derivatives. Motivated by a recent paper of Korpas and Manschot
\cite{Korpas:2017qdo} we make some comments on the extent to which one can write
the $u$-plane integrand as a total derivative in Appendix \ref{app:TotalDerivative}.

While we find the physical argument we will give compelling our proposal is
nevertheless conjectural. We can only offer some fairly limited evidence that it
is correct:  First, we remind the reader that - as already observed long ago in
 \cite{Marino:1998tb,Marino:1998uy} - the very existence of the twisted partition function in the limit $m\to m_*$ is rather
 nontrivial.   The most striking new
piece of evidence is that the correlation functions satisfy the
$U(1)_R$-charge selection rules expected for the topologically twisted $AD3$
theory on $X$ \cite{Gukov:2017zao,Marino:1998bm,Shapere:2008zf}. However, since the
background charge was in fact determined from the behavior of the $u$-plane measure
in the first place this is not really a very strong piece of evidence.
\footnote{As we review in section \ref{sec:U1R-Anomaly} below, the
 discussion of Shapere and Tachikawa \cite{Shapere:2008zf} expresses
the  $U(1)_R$  background charge in terms of the
conformal anomaly coefficients $a,c$. These authors then derived the values
of $a,c$ for the $AD3$ theory from the anomalous behavior of the $u$-plane measure. Although
we do not directly use the $U(1)_R$ anomaly in our derivation of the $AD3$ correlators, our
selection rule can be traced to the $U(1)_R$ depenence of the $u$-plane
measure,  so it is hardly
surprising that if we input the Shapere-Tachikawa values of $a,c$ into the
general formula for the $U(1)_R$ background charge we rederive
our selection  rule!   Fortunately, there are some other discussions  of these
coefficients \cite{Aharony:2007dj,Shimizu:2017kzs,Xie:2013jc} which are asserted to be
logically independent of the computation of Shapere and Tachikawa.}
%
%
In section \ref{sec:Continuous} we give very
explicit formulae for the continuous metric dependence when $b_2^+(X)=1$.  If it could be checked
more directly that would be very helpful. Indeed, any direct checks of (or counterarguments against!)
our proposal would be most welcome.

Even within the extremely limited context of the correlators for twisted AD3
theory we have left many unanswered questions. The extension to manifolds
with $b_1\not=0$ is of some interest for two reasons. First, in this case
the 3-form descendent of the $0$-observable has negative ghost number
and hence the ghost number selection rule admits the possibility that
there is an infinite number of nonzero correlation functions, in strong
contrast to the simply-connected case.
Moreover, non-simply connected manifolds are probably best suited for
comparison with the approach to computing topologically twisted d=4 N=2
partition functions suggested in \cite{Gukov:2017zao}. The extension of
our computations (and indeed of the original computations in \cite{Moore:1997pc}) to the
case $b_2^+=0$ should be quite interesting. We must note that there is some tension between
our conjecture and some remarks in \cite{Dedushenko:2017tdw} (located between
their equations $(5.14)$ and $(5.16)$) so further study of the $b_2^+=0$ case
is called for.  Finally, the microscopic
interpretation of our partition function in terms of moduli spaces of traditional
partial differential equations is an interesting open problem.
In principle one should be able to translate our definition \eqref{eq:ZAD-Def} into
some subtle aspect of intersection theory on the moduli space of the
nonabelian monopole equations, but we suspect there is a more compelling
formulation.

\section*{Acknowledgements}

We thank C. Cordova, P. Feehan, J. Harvey, G. Korpas, P. Kronheimer, C. LeBrun, J. Manschot, H. Nakajima,
W. Peelaers, Y. Tachikawa, E. Witten, and C. Woodward for useful discussions and
correspondence. A preliminary version of these results was presented by GM at the SCGP
Workshop: Gauge Theory and Low Dimensional Topology, April 26, 2017. GM thanks the organizers
S. Donaldson, K. Fukaya, and J. Morgan for the inviation to speak.
G.M. and I.N.  are supported by the U.S. Department of Energy under grant DOE-SC0010008.

\section{$SU(2)$ $N_f=1$ and AD3}\label{sec:SU2-AD3}

Here we review some well-known facts. See \cite{Tachikawa:2013kta} for a more extensive
discussion.

The Seiberg-Witten curve $\Sigma$  for $SU(2)$ theory coupled to a single hypermultiplet in the
fundamental was first presented in \cite{Seiberg:1994aj}. The class S presentation is
\cite{Gaiotto:2009hg}:
\be\label{eq:SU2Nf1}
\lambda_1^2 = \left(\frac{\Lambda_1^2}{z} + 3u + 2\Lambda_1 m z +  \Lambda_1^2 z^2  \right) \left( \frac{dz}{z} \right)^2
\ee
where $\Lambda_1$ is the UV scale, $m$ is the mass of the hypermultiplet, $u$ is a coordinate on the
Coulomb branch,  and $z \in C \cong \IC^*$ is a
coordinate on the UV curve.    The Seiberg-Witten curve is a subset of  $T^* C $ where the restriction of the
canonical Liouville one-form on $T^*C$ to $\Sigma$ is the canonical Seiberg-Witten differential.

As observed in \cite{Argyres:1995xn} when $m = \frac{3}{2} \omega \Lambda_1$, with $\omega$ a third root of
unity, three branch points of the curve collide and the discriminant of the curve has a multiple zero.
For definiteness we consider the limiting behavior as $m \to m_*:=  \frac{3}{2} \Lambda_1$, so the
discriminant has a double zero at $u=u_*:= \Lambda_1^2$ where two roots $u_\pm(m)$ collide.
To define a scaling limit we change variables:
\be
m = \frac{3}{2} \Lambda_1 + \delta m \qquad u = \Lambda_1^2 + \delta u \qquad z = -1 + \tilde z
\ee
define
\be\label{eq:Nf1-to-AD3}
\begin{split}
\tilde z & = - \epsilon z_{AD} \\
4\Lambda_1 \delta m - 3 \delta u & = 3 \epsilon^2 \Lambda_{AD}^2 \\
2 \Lambda_1 \delta m - 3 \delta u & = - \epsilon^3 u_{AD} \\
\lambda & = \epsilon^{5/2} \lambda_{AD} \\
\end{split}
\ee
and take $\epsilon \to 0$ holding all quantities with subscript $AD$ fixed. The result is
the AD3 family of curves as used in \cite{Gaiotto:2009hg}:
\be\label{eq:AD-Family}
\lambda^2 = (z^3 - 3 \Lambda^2 z + u ) (dz)^2 ~ .
\ee
Note that in the above equation we have dropped the subscript $AD$ on $\lambda,z,\Lambda,u$ to avoid
clutter. We will continue to do this in what follows, and we trust which $u$-plane is meant
will be clear from context. We sometimes write $\Lambda_{AD}$ when we wish to emphasize that it is
the mass deformation in the $AD3$ theory but often we simply write this as $\Lambda$ to avoid clutter.
The superconformal point is described by the limit $\Lambda_{AD}\to 0$ at the origin
of the $AD3$ Coulomb branch $u=0$.

A key point made in  \cite{Argyres:1995xn} is that  at
the points $u_\pm(m)$ of the $SU(2)$, $N_f=1$ family,  hypermultiplets with mutually
nonlocal charges become massless.   Therefore, when $m\to m_*$ and $u\to u_*$ there are
massless nonlocally related particles and the
low energy effective theory cannot be a Lagrangian field theory. It is, in fact,
the $AD3$ theory weakly coupled to other degrees of freedom in the $SU(2)$, $N_f=1$ theory.

We remark that the $AD3$ theory was first discovered at a point in the Coulomb branch
of pure $SU(3)$ $N=2$ SYM \cite{Argyres:1995jj}. However, in that case the $U(1)$
flavor symmetry associated with the mass parameter is gauged and therefore integrated
over. For our purposes it is much better to keep it as a free parameter.

\section{$u$-Plane Integrals}

A systematic derivation of the Witten conjecture of four-manifold theory
(equation $(2.14)$ of \cite{Witten:1994cg})
relating the Donaldson and Seiberg-Witten invariants was presented in
\cite{Moore:1997pc}. It involves an integral over the Coulomb branch of the
$SU(2)$ $N_f=0$ theory.
\footnote{Mathematically rigorous proofs of the Witten conjecture have been
given in \cite{GNY} for complex algebraic manifolds and in \cite{FeehanLeness}
for all standard four-manifolds of Seiberg-Witten simple type.}
It is referred to informally as the ``$u$-plane integral.''
For additional background and discussion of the $u$-plane integral see
 \cite{Korpas:2017qdo,Labastida:2005zz,MooreLectureNotes}.
The original discussion of \cite{Moore:1997pc} applied just to $SU(2)$
Yang-Mills coupled to $N_f \leq 4$ fundamental hypermultiplets or one
adjoint hypermultiplet, but in fact the measure makes sense for any one-dimensional
Coulomb branch.
\footnote{What is far less obvious is whether the measure is single-valued
on the $u$-plane and whether the integral over the $u$-plane is well-defined
for other families of Seiberg-Witten curve and differential. }
Although the integral is, conceptually, best written as an integral over
the $u$-plane, the path integral derivation leads more naturally to an integral over
a special coordinate $a$ so that it becomes
\begin{equation}\label{eq:Zu-DEF}
Z_u = K_u e^{2\pi \I \lambda_{0,\infty}^2}
\int da d \bar a A^\chi B^\sigma  e^{2 p u + S^2 T(a)} \Psi
\end{equation}
\begin{equation}
\Psi := \sum_{\lambda\in \lambda_{0,\infty} + \Gamma} e^{2\pi \I (\lambda-\lambda_{0,\infty} )\cdot \xi_\infty} \CN_\lambda
\end{equation}
\be
\CN_\lambda :=\frac{d \bar \tau}{d \bar a}
\frac{e^{\frac{S_+^2}{8\pi y} \left( \frac{du}{da}\right)^2} }{\sqrt{y}}
   e^{-i \pi \bar \tau \lambda_+^2 - i \pi \tau \lambda_-^2 - i \frac{du}{da} S\cdot \lambda_- }
\left[ \lambda_+ + \frac{i}{4\pi y} \frac{du}{da} S_+  \right]
\ee

Our notation is the following:

\begin{enumerate}

\item $p$ is a fugacity conjugate to the insertion of the $0$-observable $\CO=u$ in the twisted partition function.
$S\in H_2(X;\IZ)$ is a homology class and determines a canonical $2$-observable   $\CO(S) :=\int_{S} K^2 u$,
via the descent formalism \cite{Moore:1997pc}. Here $K$ is a one-form supercharge such that $[K,\CQ]=d$.
The expression $Z_u$ should be viewed as a formal
power series in $p,S$ and it is the contribution of the Coulomb branch to the correlation function
\be
\langle e^{p\CO + \CO(S) } \rangle
\ee
in the twisted theory on $X$.

\item The measure factors are
\be
A : = \alpha \left( \frac{du}{da}\right)^{1/2} \qquad  B:= \beta \Delta^{1/8}
\ee
and correspond to the terms in the low energy effective action on the Coulomb branch
describing the coupling of the $U(1)$ vectormultiplet to the Euler character $\chi$ and
the signature $\sigma$. Here $\Delta = \prod_s (u-u_s)$ is a holomorphic function
with first order zeroes at the discriminant locus $\{ u_s \}$ where a hypermultiplet becomes
massless.
\footnote{As pointed out in \cite{Shapere:2008zf} one should, in general, distinguish the ``physical
discriminant'' from the ``mathematical discriminant.'' For our main example the two will be
equal up to a constant.}
The factors  $\alpha, \beta$ are independent of $u$ but
can depend on the theory, the scale $\Lambda$, and the masses. In principle they could
also vary nontrivially on the
the conformal manifold in the superconformal case. For example, detailed analysis of the
mass-deformed $N=2^*$ theory strongly suggests that they depend on $\tau_0$ in that case
\cite{Labastida:1998sk}.
 It would be very interesting to clarify this last point and understand their dependence
 on the conformal manifold in general class S theories.

\item $a$ is a special coordinate suitable to a duality frame at  $u \to \infty$. It is the
period of the Seiberg-Witten differential on a cycle that is invariant (up to a sign) under
the path $u\to e^{2\pi \I } u$ at large $\vert u\vert$. Once we
choose a B-cycle we have $\tau = x + \I y$, decomposed in terms of real and imaginary parts. In the case of
$SU(2)$ with $N_f < 4$ this is a frame in which   $y = \Im \tau \to \infty$ as $u\to \infty$.

\item The lattice $\Gamma := H^2(X;\IZ)/{\rm Tors}$, where ${\rm Tors}$ is the
torsion subgroup of $ H^2(X;\IZ)$. The sum $\Psi$ is, essentially, the
classical partition function of the $U(1)$ gauge field on the four-manifold.
We think of $\Gamma$ as embedded in the quadratic vector space
$H^2(X;\IR)$. We have introduced a shift $\lambda_{0,\infty}$ and a phase $\xi_{\infty}$. In the case of $SU(2)$, $N_f=0$
we have $\lambda_{0,\infty}= \half \overline{w_2(P)}$ where $P$ is a principal $SO(3)$ bundle,  and $\xi_{\infty} =  \half \overline{w_2(X)}$ and the overline denotes an integral lift.
When $\tau\to \infty$ as $u\to \infty$,   $2\xi$ must be a characteristic vector on $\Gamma$ for the measure to be well-defined.
In this case we can write
\footnote{Note that general vectors in the torsor $\lambda_{0,\infty} + \Gamma$ are denoted by $\lambda$.
There is an unfortunate clash of notation with the standard notation for the Seiberg-Witten differential.
Which one is meant should be clear from context. We will mostly be using $\lambda$ to denote a vector in
the cohomology torsor from now on.}
 $\lambda = v + \lambda_{0,\infty}$, $v\in \Gamma$ and the phase
\be
e^{2\pi \I (\lambda-\lambda_{0,\infty})\cdot \xi} = (-1)^{v\cdot w_2(X)}
\ee
In the $SU(2)$ $N_f>0$ case we must take $w_2(P) = w_2(X)$ so we should take $2\lambda_{0,\infty}= 2\xi_{0,\infty}$ to be
an integral lift of $w_2(X)$.

\item  The $u$-plane integrand depends on a choice of Riemannian metric on $X$, but the dependence
only enters through the cohomology class of a self-dual two-form   $\omega \in H^2(X,\IR)$, so
$\omega = * \omega$. We can normalize it such that   $\int_X  \omega^2 = 1$. When $b_2^+ = 1$
there is a Lorentzian signature on the quadratic space $H^2(X,\IR)$,  and we must choose a component of the lightcone,
which we can call the ``forward light cone,'' in order to specify $\omega$ uniquely. Such an $\omega$ is sometimes
called a \emph{period point}. In the path integral derivation
of the $u$-plane integral one must integrate over the fermion zeromodes
and this requires a choice of orientation of   the vector space $H^0(X;\IR) \oplus H^1(X;\IR) \oplus H^{2,+}(X;\IR)$.
For $SU(2)$, $N_f=0$ this corresponds
nicely to Donaldson's discussion of orientations of instanton moduli space \cite{DK}. For $H^1=0$ and $b_2^+=1$
such an orientation  amounts to a choice of ``forward'' component of the light-cone.
Finally, we define $\lambda_+ := \lambda\cdot \omega$ and $\lambda_- := \lambda - \lambda_+ \omega$.

\item $T(a)$ depends on the choice of duality frame and is known as a ``contact term.'' It is given by
\be
T(a) = - \frac{1}{24} E_2(\tau)\left( \frac{du}{da} \right)^2 + H(u)
\ee
where it is claimed in \cite{Moore:1997pc} that $H(u) = u/3$ for all
$N_f < 4$.   For systematic treatments of such contact terms in
twisted four-dimensional N=2 theories see \cite{Edelstein:2000aj,Labastida:2005zz,Losev:1997tp}.
 Several arguments show that for the
$AD3$ theory $H(u)=0$.

\item All anti-holomorphic dependence of the integrand on $\bar u$, and all metric dependence of the integrand is
subsumed in the expression $\CN_\lambda$.

\item $K_u$ is a numerical normalization.
In TFT path integrals carry a canonical normalization: They count solutions to equations.
However, the correct normalization factor for the $u$-plane integral is not obvious. Note,
for example that there is a extra factor of $e^{2\pi\I \lambda_{0,\infty}^2}$ which could have been
absorbed into $K_u$, but we leave it this way because  a shift of $\lambda_{0,\infty}$ corresponds to a change of
sign of Donaldson polynomials familiar from Donaldson theory.
Note that for $b_2^+=1$ and $b_1=0$ we have $\chi + \sigma=4$. Therefore there is some
ambiguity in how we normalize $K_u, \alpha, \beta$ since
\be\label{eq:chgeconst}
K_u \alpha^\chi \beta^\sigma  = (K_u \kappa^{-4}) (\kappa \alpha)^\chi (\kappa\beta)^\sigma
\ee
for any nonzero constant $\kappa$.
Thus we must regard the normalization constants
\be
(K_u, \alpha, \beta) \sim (K_u \kappa^{-4}, \kappa\alpha, \kappa\beta )
\ee
as equivalent.

\item The $u$-plane integral is quite subtle and requires careful definition.
The integrand is typically quite singular at points $u=u_s$ in the $u$-plane
corresponding to zeroes of the discriminant. The procedure for defining the
integral, outlined in \cite{Moore:1997pc}, is to cut out a small disk around $u_s$,
perform the angular integral and then take the radius to zero.

\item For some four-manifolds $X$ it is possible to write the integrand of the $u$-plane integral as a
total derivative on the Coulomb branch and evaluate the integral as a sum of contours
around the singular points. This kind of representation will be important to our extraction
of the AD3 contributions to the $u$-plane integral. For details see Appendix \ref{app:TotalDerivative}.

\item We have written the $u$-plane integral in a form that applies to any one-dimensional
Coulomb branch. In this paper we will apply it to the $SU(2)$, $N_f=1$ family
\eqref{eq:SU2Nf1} and the $AD3$ family \eqref{eq:AD-Family}, in which case we will
write $Z_u^{SU(2),N_f=1}$ and $Z_u^{AD Family}$, respectively.  As mentioned above, it
would be quite
interesting to investigate the integral for the other one-dimensional Coulomb branches
described in \cite{Argyres:2015ffa,Argyres:2015gha,Argyres:2016xmc}.

\end{enumerate}

\section{The Topological Partition Function}

The full partition function of the topologically twisted theory on $X$:
\be
Z =  \langle e^{p  u + \CO(S) } \rangle,
\ee
where $\CO(S) = \int_S K^2 u$ is the canonical 2-observable associated to $S$,
is a sum of the $u$-plane integral together with contributions that guarantee that the
contribution of the vacua near $u\cong u_s$ gives a topologically invariant answer -
up to known metric dependence from the region $u\to \infty$.
It is logically possible that in theories other than $SU(2)$ coupled to matter,
new four manifold invariants other than the Seiberg-Witten invariants
(but with exactly the same wall-crossing behavior when $b_2^+(X)=1$)   can be used to achieve this
topological invariance. However, especially for theories such as the $AD3$ theory
which appear in the IR limit of Lagrangian theories, we find this
exceedingly unlikely. In any case, proceeding using the basic logic of
\cite{Moore:1997pc} the Seiberg-Witten invariants are sufficient to do the job.
The full partition function can therefore be written as:
\be
Z =  Z_u + Z_{SW}
\ee
where
\be
Z_{SW} =  \sum_s Z(u_s)
\ee
and the sum over $s$ is a sum over the discriminant locus of the family of
Seiberg-Witten curves. When the family of elliptic curves in a neighborhood of
$u_s$ is of Kodaira type $I_1$ (i.e.,  the discriminant has a first order zero at
$u=u_s$ while the Weierstrass invariants $g_2,g_3$ are nonzero at $u_s$)
the  method used in \cite{Moore:1997pc} can be applied to derive
\be\label{eq:SW-s-Exp}
\begin{split}
Z(u_s) & = \left( \sqrt{32}\pi K_u \beta^{\sigma} \alpha^{\chi} \right)
\cdot \left( e^{2\pi \I (\lambda_{0,\infty}^2 - \xi_s \cdot \lambda_{0,s} )} \eta_s \right) \\
& \sum_{\lambda\in \lambda_{0,s} + \Gamma } e^{2\pi \I \lambda\cdot \xi_s }(-1)^{n(\lambda)} {\rm SW}(\lambda) \cdot  \\
& \left[ \left(\frac{a_s}{q_s}\right)^{\chi_h-1} \frac{du}{dq_s} \left( \frac{\Delta}{q_s} \right)^{\sigma/8}
\left( \frac{da_s}{du}\right)^{1-\chi/2} e^{2pu + S^2 T_s(a_s) - \I \lambda \cdot S \frac{du}{da_s} } q_s^{-n(\lambda)} \right]_{q_s^0} \\
\end{split}
\ee
Here $\lambda_{0,s}$ and $\xi_s$ are the theta characteristics resulting from the duality transformation applied to $\Psi$
in the neighborhood of $u_s$. Similarly, $\eta_s$ is a root of unity arising from the multiplier system in that
duality transformation. The expression only makes sense for
 $\lambda_{0,s} = \half \overline{w_2(X)}$ so that the sum on $\lambda$ can be interpreted
as a sum over the characteristic class of spin-c structures on $X$.
\footnote{
$(2\lambda)$ is the characteristic class of the spin-c structure, modulo torsion.}
Then ${\rm SW}(\lambda)$ is the corresponding Seiberg-Witten
invariant associated with the Seiberg-Witten moduli space of real dimension $2n(\lambda)$ where
\be
n(\lambda) = \half \lambda^2 - \frac{\sigma}{8}  - \chi_h
\ee
and $\chi_h:= (\chi + \sigma)/4$. (For a complex surface $\chi_h$ is the holomorphic Euler characteristic.)
The special coordinate $a_s$ vanishes at $u_s$ and the coordinate $q_s = e^{2\pi \I \tau_s} \to 0$
as $u \to u_s$. The contact term $T_s(a_s)$ is obtained from $T(a)$ by duality transformation.

\section{Deriving The AD3 Partition Function}\label{sec:Deriving}

In order to extract the partition function of the AD3 theory from that of
the $SU(2)$, $N_f=1$ theory we use the following principles:

\begin{enumerate}

\item The limit of $Z^{SU(2), N_f=1}$ as $m\to m_*$ must exist since there are no noncompact Higgs branches.
(Noncompact Higgs branches are the only source of IR divergences given that $X$ is compact and the
contribution from $u\to \infty$ is finite.)

\item The resulting path integral must be an integral over all $Q$-invariant field configurations.

\item According to  \cite{Argyres:1995xn}  those $Q$-invariant configurations include the supersymmetric ``states" of the AD theory,
perhaps coupled to other degrees of freedom in the $SU(2)$ theory. However, at $m=m_*$ those couplings should be arbitrarily
weak in the scaling region of the $u$-plane near $u_*$.

\item We can therefore isolate the AD configurations by focusing on the contribution from an infinitesimally small neighborhood
of the colliding singularities $u_\pm(m)$ plus the SW contributions associated with just those points.

\end{enumerate}

When $m=m_*$, the family \eqref{eq:SU2Nf1} has a singularity at   $u=u_*$, where two
singularities $u_\pm(m)$ have collided, and another singularity $u_0$ far away from the scaling region.
Since the definition of the integral requires a subtle regularization over the noncompact regions
it turns out that:
\be\label{eq:LimInt-ne-IntLim}
  Z_u^{SU(2), N_f=1} -    \int du d \bar u \lim_{m \to m_*}  \left( \vert \frac{da}{du} \vert^2
A^\chi B^\sigma e^{2pu + S^2 T(u)} \Psi \right)
\ee
has a nonzero Laurent expansion in powers of $\mu^{1/4}$ around $\mu=0$ where $\mu:= (m-m_*)/\Lambda_1$.
Here the integral of the $m\to m_*$ limit of the integrand of $Z_u^{SU(2), N_f=1}$    is defined by cutting out disks around $u_0$ and $u_*$ and taking the limit as the disks shrink. The singular terms in the expansion \eqref{eq:LimInt-ne-IntLim} will cancel against
similar singular terms from $Z_{SW}$. The constant term (i.e. the coefficient of $\mu^0$)
is in general nonzero and does not cancel against the
constant term from $Z_{SW}$.

The quantity \eqref{eq:LimInt-ne-IntLim} comes from the integration around
an infinitesimal region near $u=u_*$. Indeed, for any $\epsilon>0$ let $B(\epsilon;u_*)$ be a disk around
$u_*$ with $\vert u - u_* \vert < \epsilon$. When $m$ is sufficiently close to $m_*$ the two colliding singularities
$u_\pm(m)$ will be inside this disk. Therefore, for any fixed $\epsilon>0$:
\be\label{eq:CommLim}
\begin{split}
\lim_{m \to m_*}  \int_{\IC-B(\epsilon;u_*)}  du d \bar u   \left( \vert \frac{da}{du} \vert^2 A^\chi B^\sigma e^{2pu + S^2 T(u)} \Psi
\right)
 & =  \\
  \int_{\IC-B(\epsilon;u_*)}  du d \bar u \lim_{m \to m_*} &
 \left( \vert \frac{da}{du} \vert^2A^\chi B^\sigma e^{2pu + S^2 T(u)} \Psi \right)\\
\end{split}
\ee
Therefore, in view of the limiting behavior reviewed in section \ref{sec:SU2-AD3} we should
 attribute the difference \eqref{eq:LimInt-ne-IntLim} to  the contribution of the AD partition function
on four-manifolds with $b_2^+ =1 $.

In the $SU(2)$ $N_f=1$ family when $m \to m_*$ and $u\to u_*$ the AD3 theory is still weakly coupled
to other degrees of freedom in the original gauge theory.
The detailed considerations of Appendix \ref{app:TotalDerivative} and Appendix \ref{app:Zutilde}
show that we should extract a factor  $\exp[ 2p\left(u_*+ \frac{2}{3}\mu\right) +  S^2 T_*]$
to account for these couplings. Here and henceforth we will choose units so that $\Lambda_1=1$.
The peculiar shift by $2\mu/3$ in the coefficient of $p$ is due to the linear combinations
 \eqref{eq:Nf1-to-AD3}. Thus we consider the constant term in the Laurent expansion around $\mu=0$:
\be\label{eq:Extract}
\Biggl[e^{-2p\left(u_*+ \frac{2}{3}\mu\right) - S^2 T_*}\left(
   Z_u^{SU(2), N_f=1}  -    \int du d \bar u \lim_{m \to m_*}  \left( A^\chi B^\sigma e^{2pu + S^2 T(u)} \Psi \right) \right) \Biggr]_{\mu^0}
\ee
Again, the detailed considerations of Appendix \ref{app:TotalDerivative} and Appendix \ref{app:Zutilde}
strongly motivate the following  conjectures:

\begin{enumerate}

\item The constant term in \eqref{eq:Extract} is in fact a \underline{polynomial}
in $p$ and $S$,  in striking contrast to the partition functions of Donaldson-Witten theory.
   We will denote it by $P_1(p,S)$.

\item
Furthermore,  if we define a grading of the polynomial $P_1(p,S)$ by ``$R$ charge'' with
 $R[p]= 6$ and $R[S]=1$ then   the highest degree is given by $6\ell + r = \fB := -\frac{1}{4}(7\chi + 11 \sigma)$.

\item  If one considers the $u$-plane integral for the $AD3$ family \eqref{eq:AD-Family} it has a similar
expansion in powers of $\Lambda_{AD}^{1/2}$ around $\Lambda_{AD}=0$ and the constant term
 $P_{AD}(p,S)$ is also a polynomial in $p$ and $S$.

\item
 Finally, defining $P_1^{\rm top}(p,S)$ be be the sum of terms with maximal
 $R$-charge we have:
\be\label{eq:Rescalings}
P_1^{\rm top}(p,S) = N P_{AD}(n_0 p, n_2 S)
\ee
for suitable constants $N,n_0,n_2$.~
\footnote{We interpret the  terms of lower $R$-charge in the polynomial $P_1(p,S)$ as effects arising from the coupling of the
AD3 theory to other degrees of freedom in the $SU(2)$ $N_f=1$ theory. It would certainly be useful to understand the physics of the lower order terms better.}

\end{enumerate}

 The results of Appendix \ref{app:TotalDerivative} and Appendix \ref{app:Zutilde}
are enough to prove all the above claims for the \underline{difference} of $u$-plane integrals for any two choices of metric.
Moreover, they   establish the above claims absolutely when $X$ has a homotopy type so that the $u$-plane integral
has a vanishing chamber in the sense explained in section 5  of \cite{Moore:1997pc}.

These considerations motivate our central formula for how to extract the physics of the $AD3$ theory
from the expansion around $\mu=0$ of the $SU(2)$, $N_f=1$ partition function:
\be\label{eq:ZAD-Def}
\begin{split}
\widetilde Z_{AD} &  :=  \Biggl[e^{-2p\left(u_*+ \frac{2}{3}\mu\right) - S^2 T_*}\left(
   Z_u^{SU(2), N_f=1}  -    \int du d \bar u \lim_{m \to m_*}  \left( A^\chi B^\sigma e^{2pu + S^2 T(u)} \Psi \right) \right) \Biggr]_{\mu^0}^{\rm top} \\
&
+ \Biggl[ [e^{-2p\left(u_*+ \frac{2}{3}\mu\right)- S^2 T_*}\left(   Z_{SW}^{SU(2), N_f=1}(u_+(m)) + Z_{SW}^{SU(2), N_f=1}(u_-(m))  \right)\Biggr]_{\mu^0}^{\rm top} \\
\end{split}
\ee

On the other hand, a very natural way to define the partition function of the $AD3$ theory
is to use directly the family of curves \eqref{eq:AD-Family} and define:
\be\label{eq:ZAD-from-AD}
Z_{AD} :=  \lim_{\Lambda_{AD}\to 0} \Biggl[  Z_{u}^{AD family} + Z_{SW}^{AD family} \Biggr]
\ee
We conjecture that, up to an overall constant and a renormalization of
$p$ and $S$ as in \eqref{eq:Rescalings}, we have $\widetilde Z_{AD}  =  Z_{AD}$.  Again,
a full proof of this statement follows from the considerations of
Appendix \ref{app:TotalDerivative} and Appendix \ref{app:Zutilde}, if
we consider the difference of partition functions for two metrics, or
if we consider a homotopy type of $X$  admitting a vanishing chamber.
Moreover, if $b_2^+>1$ then the statement is an easy consequence of the
relationship of the two curves described in section \ref{sec:SU2-AD3}.

Our main conjecture is that
\be\label{eq:MainConjecture}
Z_{AD} = \langle e^{p \CO + \CO(S) } \rangle
\ee
for the topologically twisted $AD3$ theory on four-manifolds $X$ with $b_2^+>0$.

\section{The SW Contribution To $Z_{AD}$ }

When $X$ has $b_2^+>1$ only $Z_{SW}$ contributes to the partition
function. In this section we will evaluate it fairly explicitly
for the AD3 family \eqref{eq:AD-Family} in the limit $\Lambda_{AD}\to 0$.
Thus we are starting from the definition \eqref{eq:ZAD-from-AD}.

\subsection{A General Simplification Of $Z(u_{s})$}

To begin we put \eqref{eq:SW-s-Exp} in a form which is more suitable for
explicit evaluation. In fact our derivation of the result \eqref{eq:SW-s-Exp2} below applies
to any family of elliptic Seiberg-Witten curves with a simple zero of
the discriminant at $u=u_s$ such that the Weierstrass invariants $g_2,g_3$
are nonzero at $u=u_s$. (This is Kodaira type $I_1$.)
We also assume $\lambda_{0,s} = \xi_{s} = \half \overline{w_2(X)} $. This holds for the
$SU(2)$, $N_f=1$ family and therefore for the AD3 family. Moreover, the duality
transformations needed to transform from the duality frame at $u=\infty$ to $u$ near
$u_s$ are all, according to equation (11.17) of \cite{Seiberg:1994aj}, conjugate to $T$.
It turns out that the measure of the $u$-plane transforms by a character under
$S$ and $T$. Therefore the root of unity $\eta_s$ is independent of $s$ and we will just
denote it by $\eta$.

Now we can replace the sum over $\lambda$ by the average over $\lambda$ and $-\lambda$.
Because $\lambda_{0,s}=\xi_s = \half \overline{ w_2(X)}  $ we have
\be
 e^{-4\pi \I \lambda \cdot \xi_s } = e^{-2\pi \I (v+\half w_2)\cdot w_2} = e^{-\I \pi w_2^2} = (-1)^\sigma.
\ee
Moreover it is a standard result of Seiberg-Witten theory that
\be
{\rm SW}(-\lambda) = (-1)^{\chi_h} SW(\lambda)
\ee
so in the sum over $\lambda$ in \eqref{eq:SW-s-Exp} we can freely make the replacement:
\be\label{eq:SW-average}
  {\rm SW}(\lambda) e^{2\pi \I \lambda\cdot \lambda_0}e^{-\I \lambda \cdot S \frac{du}{da_s} } \rightarrow
  \half {\rm SW}(\lambda) e^{2\pi \I \lambda\cdot \lambda_0}
\left( e^{-\I \left( \frac{d u}{da_s} \right) S\cdot \lambda } + (-1)^{\chi_h + \sigma} e^{\I \left( \frac{d u}{da_s} \right) S\cdot \lambda } \right)
\ee
The reason this is useful is that the expansion in $S\cdot\lambda$ only involves powers of   $\left( \frac{d a_s}{du} \right)$
of a definite parity independent of $\lambda$.
That will be important since, as we will see below, we can readily determine the $q_s$-expansion of
$\left( \frac{d a_s}{du} \right)^2$ near $u_s$, but taking the square-root could be tricky.
Equation \eqref{eq:SW-average} motivates us to define:
\be
\half
\left( e^{-\I \left( \frac{d u}{da_s} \right) S\cdot \lambda } + (-1)^{\chi_h + \sigma} e^{\I \left( \frac{d u}{da_s} \right) S\cdot \lambda } \right) :=
\sum_{n\geq 0} \hat{c}_n^{\chi_h+ \sigma}(S)  \left( \frac{d a_s}{du} \right)^{ - n}
\ee
with
\be
\hat{c}_n^{\chi_h + \sigma}(S)  = \begin{cases}
\frac{e^{-\I \pi n/2}  }{n!}   (S\cdot \lambda)^n & n =  (\chi_h + \sigma)  ~\mod 2 \\
0  & n \not= (\chi_h + \sigma) ~ \mod 2  \\
\end{cases}
\ee

Now suppose we have a SW curve presented in the form:
\be\label{eq:EllipticCurve}
y^2 = x^3 + A_2 x^2 + A_4 x + A_6
\ee
and there is a special coordinate $a_s$ so that $a_s \to 0$ but
\be
\frac{da_s}{du} = \frac{\rho}{\pi} \omega_1
\ee
is nonvanishing as $q_s=e^{2\pi\I \tau_s} \to 0$.
\footnote{Here $\rho$ is a relative normalization between the standard periods $\omega$ of the
elliptic curve and $\frac{da_s}{du}$. Its value depends on the conventions used to normalize the
central charge. In the conventions of
\cite{Gaiotto:2009hg} the central charge is $Z(\gamma) = \pi^{-1} \oint_{\gamma} \lambda$,
so $\frac{da}{du} = \pi^{-1} \oint_{A} \frac{d \lambda}{d u} $.  Next, for an elliptic
curve presented in the form \eqref{eq:EllipticCurve} the canonically normalized holomorphic
differential is $\sqrt{2}\frac{dx}{y}$. Finally, we note that for the AD3 family \eqref{eq:AD-Family}
we have $\frac{d\lambda}{du} = \half \frac{dz}{y}$. We thus conclude that for natural conventions
for class $S$ we have $\rho = 1/\sqrt{8}$. However, we leave $\rho$ undetermined above since it is
different if one uses other conventions such as those of \cite{Seiberg:1994aj} or \cite{Moore:1997pc}.
The results for different choices of $\rho$ are simply related by a renormalization of $S$. }

In order to evaluate \eqref{eq:SW-s-Exp} we need to know the expansions
\be\label{eq:u-expansion}
u = u_s + \mu_1 q_s + \mu_2 q_s^2 + \cdots
\ee
\be\label{eq:as-expansion}
a_s = \kappa_1 q_s + \kappa_2 q_s^2 + \cdots
\ee
We now show how to extract these expansions - in principle - from the SW curve.

From $A_2, A_4,A_6$ we can construct the standard Weierstrass invariants $g_2,g_3$.
For $SU(2)$ theories and the AD3 family these will be polynomials in $u$. In general we have
\be
 (12)^3 \frac{g_2^3}{g_2^3 - 27 g_3^2}= j(\tau_s) = q_s^{-1} + 744 + 196884 q_s + 21493760 q_s^2 +  \cdots
\ee
Actually, for our purposes, this equation is more usefully written as
\be\label{eq:E6E4g3g2}
(27) \frac{g_3^2}{g_2^3} = \frac{E_6^2}{E_4^3}
\ee
Plugging  \eqref{eq:u-expansion} into either version gives a triangular system of equations from which
we can extract the coefficients $\mu_n$. Next, if we have chosen a basis so that $\tau = \omega_2/\omega_1$
then the period $\omega_1$ is expressed in terms of coefficients of the elliptic curve and $\tau$ by
\be
\omega_1^2 = 2 \left( \frac{\pi}{3} \right)^2 \frac{E_6(\tau)}{E_4(\tau)}\cdot \frac{g_2}{g_3}
\ee
and hence
\be\label{eq:dadu}
\left(\frac{da_s}{du}\right)^2 = 2\left( \frac{\rho}{3}\right)^2   \frac{E_6(\tau)}{E_4(\tau)}\cdot \frac{g_2}{g_3}  ~ .
\ee
Now we use the standard expansions of $E_4, E_6$ in terms of $q_s$ and we expand the polynomials $g_2,g_3$ of $u$
around $u_s$ and use \eqref{eq:u-expansion}. This gives $\kappa_1^2$ and all the  $\kappa_n/\kappa_1$
 for $n>1$.

We also write $\Delta = \CN^u_{\rm math} \Delta^{\rm math}$
where $\Delta^{\rm math}$ is the mathematical discriminant of the elliptic curve,
\be
\Delta^{\rm math}= (e_1-e_2)^2(e_1-e_3)^2(e_2-e_3)^2= 4 (4 g_2^3 - 27 g_3^2)
= 2^{-22} \left( \frac{da_s}{du} \right)^{-12} \eta(\tau_s)^{24}
\ee
where $e_i$, $i=1,2,3$ are the roots of the cubic. Putting all these things together
we can write  \eqref{eq:SW-s-Exp} in the form
\be\label{eq:SW-s-Exp2}
\begin{split}
Z_{SW}^s & = \left( \sqrt{32}\pi K_u \beta^{\sigma}\alpha^{\chi} 2^{-11\sigma/4} (\CN^u_{\rm math})^{\sigma/8} \eta \right)
  \\
& \sum_{n\geq 0} \sum_{\lambda\in \half w_2 + \Gamma } e^{\pi \I \lambda\cdot w_2  } (-1)^{n(\lambda)} {\rm SW}(\lambda) \cdot
\hat c_n^{\chi_h + \sigma}(S)  \\
& \left[\frac{du}{dq_s} \left( \frac{\eta(\tau_s)^{24} }{q_s} \right)^{\sigma/8} \left(\frac{a_s}{q_s}\right)^{\chi_h-1}
\left( \frac{da_s}{du}\right)^{1-2\chi_h - \sigma-n } e^{2pu + S^2 T_s(a_s)}  q_s^{-n(\lambda)} \right]_{q_s^0} \\
\end{split}
\ee
This result is a slight generalization of, and improvement upon, equation $(11.28)$ of \cite{Moore:1997pc}.

\subsection{Specializing To The AD3 Family}

We now specialize \eqref{eq:SW-s-Exp2} to the $AD3$ family of curves.

For the $AD3$ family we have $g_2 = 3 \Lambda^2$ and $g_3 = - u/2$, $\Delta$ is
quadratic in $u$ and so there are just two singularities $u_\pm$.
\footnote{In this section we write $\Lambda$ instead of $\Lambda_{AD}$.}
Near
each of them we have the expansion in $q_s$:
\be
u = u_s \frac{E_6}{E_4^{3/2}}
\ee
with $u_s = 2 \zeta_s \Lambda^3$. Here $\zeta_s = \pm 1$ at the two singularities
and $E_6, E_4$ are power series in $q_s$ beginning at $1$. Fractional powers of
Eisenstein series are to be interpreted as power series in $q_s$. Note that
\be
\frac{du}{dq} = - \zeta_s (12  \Lambda)^3 \left( q^{-1} \eta^{24} \right)\cdot E_4^{-5/2}
\ee
From \eqref{eq:dadu} we obtain $\kappa_1^2 = - \half (12\Lambda )^5 \zeta_s $ and
\be
{\check E}_1(q):= \frac{a_s/\kappa_1}{q} := 1 + \sum_{n\geq 2}^\infty \frac{\kappa_n}{\kappa_1} q^{n-1}
\ee
is independent of $s$ and satisfies the equation:
\be
q \frac{d}{dq} (q {\check E}_1(q)) = \eta^{24} E_4^{-9/4} = (12)^{-3} (E_4^3-E_6^2)E_4^{-9/4}
\ee
from which one may generate its $q$-series. There does not appear to be any
simple expression for ${\check E}_1$ in terms of  $E_2$,  $E_4$ and $E_6$ and we will, regrettably,
take the above as its definition.

Using these formulae and \eqref{eq:SW-s-Exp2} one can derive
\be\label{eq:AD3-Family-BigMess}
\begin{split}
Z_{SW}^{AD3 family} = \CC_1 \sum_{r_1\geq 0} \delta_{r_1} & \sum_{\lambda} \sum_{\zeta_s=\pm 1} \Lambda^{\half ( r_1 - (\chi_h - c_1^2))}   e^{\I \pi \lambda\cdot w_2} (-1)^{n(\lambda)} {\rm SW}(\lambda) \frac{  (\sqrt{24}S\cdot \lambda)^{ r_1} }{(4\rho)^{r_1} r_1 !} \\
\zeta_s^{2\chi_h + \sigma + r_1 - \half(\chi_h + \sigma + r_1)}
& \left[ {\check E}_1^{\chi_h-1} (q^{-1} \eta^{24})^{1 + \frac{\sigma}{8}}
E_4^{-\frac{1}{4}( 2\chi_h + \sigma + 9 + r_1)}
e^{4 (\zeta_s \Lambda^3 p) E_6 E_4^{-3/2} + (\zeta_s \Lambda \left(\frac{S}{4\rho}\right)^2) E_2 E_4^{-1/2} } \right]_{q^{n(\lambda)}} \\
\end{split}
\ee
where $\delta_{r_1}$ enforces the constraint $ r_1=(\chi_h + \sigma)\mod 2$ and
%
%
%
\be
\CC_1 = \left( \sqrt{32}\pi  K_u \cdot \beta^\sigma \cdot \alpha^\chi  \eta \right)
\left(-\I 2^{11/2} 3^{7/2} \rho^{-1}\right)^{\chi_h} \left(\I 2^{-13/4} 3^{1/2} \rho^{-1} \CN^u_{\rm math} \right)^{\sigma}
\ee

Next, we expand the terms with $p$ and $S^2$ in the exponential. We find that the terms proportional to
$(S\cdot \lambda)^{r_1} (S^2)^{r_2} p^\ell$ come with the power $\Lambda^{U/2}$ where
\footnote{The quantity $\fB$ is very natural in this subject. The quantity
$2\fB$ provides a lower bound for the number of Seiberg-Witten basic
classes of $X$ \cite{Marino:1998tb}. }
\be
U:= r_1 + 2r_2 + 6 \ell - \fB \qquad\qquad \fB := \chi_h - c_1^2 = - \frac{7 \chi + 11 \sigma}{4}
\ee
This is, of course, a reflection of the emergent $U(1)_R$ symmetry at the superconformal point.

Next, the entire dependence of the expression \eqref{eq:AD3-Family-BigMess}
on the two values $s=+$ and $s=-$  is summarized by the power
\be
\zeta_s^{\half( \sigma+r_1 -\chi_h) + \ell + r_2},
\ee
so the sum over $\zeta_s$ imposes the selection rule $U= 0~ \mod ~ 4$. (This selection rule
implies that $r_1 = (\chi_h + \sigma )\mod 2$ so we can now drop that constraint.) The result of these
considerations is that:
\be
\begin{split}
Z_{SW}^{AD3 family}& =2 \CC_1 \sum_{U=0 \mod 4} \sum_{\lambda}
\Lambda^{\half U }
 e^{\I \pi \lambda\cdot w_2} (-1)^{n(\lambda)} {\rm SW}(\lambda) \frac{  (\sqrt{24}S\cdot \lambda)^{r_1} }{(4\rho)^{r_1}r_1 !}
\frac{(S^2)^{r_2}}{(4\rho)^{2r_2 }r_2!} \frac{(4p)^\ell}{\ell!}  \\
& \left[ {\check E}_1^{\chi_h-1} (q^{-1} \eta^{24})^{1 + \frac{\sigma}{8}} E_2^{r_2}
E_4^{-\frac{1}{4}(9 + U - 5 \chi_h )} E_6^\ell \right]_{q^{n(\lambda)}} \\
\end{split}
\ee
where the first sum is over all integers $r_1,r_2,\ell\geq 0$ such that $U=0\mod 4$.

Now we wish to take the $\Lambda \to 0$ limit. We can organize the sum by the degree $U$.
Note that there are potentially \underline{negative} powers of $\Lambda$ if $\fB>0$.
Nevertheless, the correlators should be finite in the $\Lambda \to 0$ limit. This was the original
argument of   \cite{Marino:1998tb} used to derive sum rules on Seiberg-Witten invariants.  However, unlike
 \cite{Marino:1998tb}, here we are not assuming that $X$ is of Seiberg-Witten simple type.
\footnote{``Seiberg-Witten simple type'' is often given the acronym SWST below.}

For any given $X$ there will be a finite number of sum rules, one for each nonnegative
integer $k$ such that $k-\fB < 0$ and $k=\fB \mod 4$. For each such $k$ the sum, for fixed degree $U = k-\fB$
must vanish. To be concrete:

\begin{enumerate}

\item Suppose $\chi_h - c_1^2 > 0$ and $\chi_h-c_1^2 = 0 \mod 4$. Then
\be\label{eq:SumRule1}
\begin{split}
0  = \sum_{\lambda}  e^{\I \pi \lambda\cdot w_2} (-1)^{n(\lambda)} {\rm SW}(\lambda)
& \left[ {\check E}_1^{\chi_h-1} (q^{-1} \eta^{24})^{1 + \frac{\sigma}{8}}
E_4^{-\frac{1}{4}(9 + c_1^2 - 6 \chi_h )}  \right]_{q^{n(\lambda)}} \\
\end{split}
\ee

\item Suppose $\chi_h - c_1^2 > 1$  and $\chi_h-c_1^2 = 1 \mod 4$. Then the $U=1-(\chi_h-c_1^2)$ only gets a
contribution from $r_1 = 1$, $r_2=\ell=0$ and hence
\be\label{eq:SumRule2}
\begin{split}
0  = \sum_{\lambda}  e^{\I \pi \lambda\cdot w_2} (-1)^{n(\lambda)} {\rm SW}(\lambda) (S\cdot \lambda)
& \left[ {\check E}_1^{\chi_h-1} (q^{-1} \eta^{24})^{1 + \frac{\sigma}{8}}
E_4^{-\frac{1}{4}(10 + c_1^2 - 6 \chi_h )}  \right]_{q^{n(\lambda)}} \\
\end{split}
\ee

\item Suppose $\chi_h - c_1^2 > 2$  and $\chi_h-c_1^2 = 2 \mod 4$. Then the $U=2-(\chi_h-c_1^2)$  gets a
contribution from $r_1 = 2$, $r_2=\ell=0$ and $r_1=0, r_2=1,\ell=0$ hence
\be\label{eq:SumRule3}
\begin{split}
0  = \sum_{\lambda}  e^{\I \pi \lambda\cdot w_2} (-1)^{n(\lambda)} {\rm SW}(\lambda) & \\
\{ S^2 \left[ {\check E}_1^{\chi_h-1} (q^{-1} \eta^{24})^{1 + \frac{\sigma}{8}}E_2
E_4^{-\frac{1}{4}(11 + c_1^2 - 6 \chi_h )}  \right]_{q^{n(\lambda)}}
& \\
- \half (-24)^{\half(\chi_h + \sigma + 1)}  (S\cdot \lambda)^2
 \left[ {\check E}_1^{\chi_h-1} (q^{-1} \eta^{24})^{1 + \frac{\sigma}{8}}
E_4^{-\frac{1}{4}(11 + c_1^2 - 6 \chi_h )}  \right]_{q^{n(\lambda)}} \} &  \\
\end{split}
\ee

\item And so on: We get rather complicated polynomials in $S^2$, and $S\cdot \lambda$ which must vanish.
If we assume SWST then only the spin-c structures with $n(\lambda)=0$ contribute and we get the criteria
of \cite{Marino:1998tb}. In this case the formulae simplify a lot because all the factors
of the form $[{\check E}_1^{\chi_h-1} \cdots ]_{q^{n(\lambda)}} $ can be put equal to $1$.

\end{enumerate}

Now we consider the actual value at $\Lambda=0$. According to our conjecture above,
this should give the partition function of topologically twisted $AD3$ theory on standard
four-manifolds. Technically, we simply keep the terms above with  $U=0$ so our formula is
\be\label{eq:AD3-StandardX}
\begin{split}
\langle e^{p\CO + \CO(S) } \rangle_X^{AD3} =2 \CC_1
 \sum_{U = 0}\sum_{\lambda}
& e^{\I \pi \lambda\cdot w_2} (-1)^{n(\lambda)} {\rm SW}(\lambda) \frac{  (\sqrt{24}S\cdot \lambda)^{r_1} }{(4\rho)^{r_1}r_1 !}
\frac{(S^2)^{r_2}}{(4\rho)^{2r_2 }r_2!} \frac{(4p)^\ell}{\ell!}  \\
& \left[ {\check E}_1^{\chi_h-1} (q^{-1} \eta^{24})^{1 + \frac{\sigma}{8}} E_2^{r_2}
E_4^{-\frac{1}{4}(9 - 5 \chi_h )} E_6^\ell \right]_{q^{n(\lambda)}} \\
\end{split}
\ee
This is the generator of correlation functions of the twisted $AD3$ theory on four-manifolds
$X$ with  $b_1=0$ and $b_2^+>1$. It is only nonvanishing for $\fB=\chi_h - c_1^2 \geq 0$.

We now assume that $X$ has Seiberg-Witten simple type (SWST) so that only spin-c structures with $n(\lambda)=0$ contribute.
Moreover, we will also assume that $X$ is  of superconformal simple type (SCST) with $\fB \geq 4$. According to
\cite{Marino:1998tb,Marino:1998uy} this means that
\be\label{eq:SCST-def}
\sum_{\lambda} e^{\I \pi \lambda\cdot w_2} {\rm SW}(\lambda) (\lambda \cdot S)^k = 0 \qquad  0 \leq k \leq \fB -4
\ee
Therefore, given the constraint $U=0$ the only terms that can contribute are $r_1 = \chi_h - c_1^2 - 2  = \fB-2$, $r_2=1$, $\ell=0$, and
$r_1 = \chi_h -c_1^2 = \fB$, $r_2=\ell=0$, and our partition function simplifies to
\be\label{eq:SimpleAD3}
\langle e^{p\CO + \CO(S) } \rangle_X^{AD3} = \CC_2 \sum_{\lambda} e^{\I \pi \lambda\cdot w_2} {\rm SW}(\lambda)
\left[ \frac{\fB(\fB-1)}{24} S^2 (S\cdot \lambda)^{\fB-2} + (S\cdot \lambda)^\fB \right]
\ee
 and to get the constant we observe
that $\Delta^{\rm math} = - 27 (u^2 - (2\Lambda^3)^2) $ so $\CN^u_{\rm math} = - 1/27$.
After some computation we find:
\be
\CC_2 = \frac{ \sqrt{128}\pi \eta'  }{\fB!} K_u \left( \frac{ 32\beta \rho^{3/2} }{3^{3/8}} \right)^{\sigma}
\left(2^{9/4} \alpha \rho^{3/2} \right)^{\chi}
\ee
where $\eta'$ is an eighth root of unity we have not determined.
(One could probably use the fact that
  $SU(2)$ $N_f=1$ theory is time-reversal invariant for $\Lambda$ real to constrain this phase.)

\subsection{Discussion: Seiberg-Witten Simple Type vs. Superconformal Simple Type}\label{subsec:Discussion}

A striking property of \eqref{eq:SimpleAD3} is that it does not depend at all on $p$.
This comes about because when the condition $U=0$ is combined with the SCST condition, the only solutions have $\ell=0$.
This means the $0$-observable is a ``null vector.'' That is, insertions of $\CO$ into correlators always
vanish for such four-manifolds. Although it is certainly true that $\CO_{\rm classical}=0$ it is not obvious
why this should be true in the quantum theory and this leads us to take seriously the possibility that
there might be standard four-manifolds that are not of SWST. Indeed, if we drop the SCST condition
\eqref{eq:SCST-def} there are many more solutions to $U=0$, i.e. $r_1 + 2 r_2 + 6 \ell = \fB$ which
will contribute to \eqref{eq:AD3-StandardX}. Some will   include $\ell\not=0$, and we cannot
use the necessary conditions \eqref{eq:SumRule1}, \eqref{eq:SumRule2}, \eqref{eq:SumRule3}, et. seq.
to eliminate the $p$-dependence.

At this point it is important to recall that reference  \cite{Marino:1998tb} derived necessary
conditions for the finiteness of the $\Lambda \to 0$ limit (these are the conditions
\eqref{eq:SumRule1}, \eqref{eq:SumRule2}, \eqref{eq:SumRule3}, et. seq.  above in the special
case of SWST). These conditions  are quite complicated so the
authors of \cite{Marino:1998tb} also formulated the  SCST condition, namely, that either $\fB \leq 3$ or
\eqref{eq:SCST-def} holds. The SCST condition is a \underline{sufficient} condition for finiteness of the $\Lambda \to 0$ limit.
The authors of   \cite{Marino:1998tb} then
 checked that all known  (as of 1998) standard four-manifolds satisfy the SCST condition and  they conjectured
that all standard four-manifolds are of SCST. The work of \cite{GNY} gave a different argument that complex algebraic manifolds are of
SCST. The work \cite{FeehanLeness} shows - subject to an unproven hypothesis - that for all standard
four-manifolds, SWST  implies SCST.  Therefore, (accepting the work of \cite{FeehanLeness}), all standard
four-manifolds of SWST have the property that the topological correlators are given by \eqref{eq:SimpleAD3},
and, in particular, the $0$-observable is  a ``null-vector.''  We reiterate that in the absence of any compelling reason for $\CO$ to be a null-vector, one must suspect that there are in fact standard four-manifolds that
are not of Seiberg-Witten simple type.

Witten has pointed out \cite{WittenPC} that the null-vector property of $\CO$ has an
interesting similarity with the appearance of the Newstead-Ramanan conjecture in the framework
of two-dimensional nonabelian gauge theory, as described in \cite{Witten:1992xu} (see section 4.3, especially
equation $(4.51)$ of that paper).

\section{The $u$-plane Contribution To $Z_{AD}$}\label{sec:Continuous}

We now turn to the $u$-plane integral $Z_u^{AD3 Family}$. We will find that, once again,
the coefficient of $\Lambda_{AD}^0$ in the expansion around $\Lambda_{AD}\to 0$  is a
polynomial with terms satisfying the selection rule $U=0$.  (In particular,
it vanishes for manifolds such as $S^2 \times S^2$ and $\IC\IP^2$, cases where the
corresponding integrals in Donaldson theory are quite interesting.)

As discussed in Appendix \ref{app:TotalDerivative} we do not know how to give a
general contour integral expression for  the  result of the $u$-plane
integral, but one key feature can be immediately noticed: In the AD3 family the
$\tau$-parameter approaches a finite value $\tau_*$ as $u\to \infty$. Just as
in case of the $SU(2)$, $N_f =4$ theory studied in \cite{Moore:1997pc} this results
in \underline{continuous metric dependence}:
The general arguments for invariance of the topological partition function fail utterly.
We expect this to be a generic feature of topologically twisted
superconformal partition functions on four-manifolds of $b_2^+=1$.

Note that for the AD3 family, even when $\Lambda_{AD}\not=0$ for $u\to \infty$ we have, in any duality frame
\be
\begin{split}
a & \rightarrow \kappa u^{5/6} + \cdots \\
a_D & \rightarrow \kappa\tau_* u^{5/6} + \cdots \\
\end{split}
\ee
where $\kappa$ is a nonzero constant and
 $\tau_*$ is in the $PSL(2,\IZ)$ orbit of $e^{\I \pi/3}$. For concreteness, we will choose a frame
so that $\tau_* = e^{\I \pi/3}$. This means that $da/du \sim u^{-1/6} + \cdots $ is not single-valued
on the $u$-plane.
It is thus quite nontrivial, and somewhat remarkable, that the $u$-plane measure is in fact well-defined
at $u \to \infty$. Nevertheless, one can  indeed check that it is well defined by directly making the modular transformation
of the integrand by $(TS)^{-1}$. From the physical viewpoint it is quite important that the measure be well-defined on
the $u$-plane and not just on some cover.

As explained in Appendix \ref{app:TotalDerivative}, it is possible to write the $u$-plane integral as a
sum of contour integrals when we consider the difference of integrals for two period points $\omega$ and $\omega_0$.
The continuous metric dependence for the $AD3$ family comes from the contour at $u\to \infty$ and, as explained in
Appendix \ref{app:TotalDerivative}, this difference can be written as $G^\omega_\infty - G^{\omega_0}_\infty$
where $G^\omega_{\infty}$ is a contour integral depending only on $\omega$ and not both $\omega,\omega_0$.
Using the expansions in \eqref{eq:E6E4g3g2-2} et. seq. we can be quite explicit. Up to an overall normalization
factor we have:
\be
\begin{split}
G_\infty^{\omega} = - \oint_{\gamma_{\infty}} \frac{du}{u} u^{-\fB/6} e^{- \frac{\textbf{w}^2}{24} E_2(\tau_*)} & \\
\Biggl\{ \sum_{\lambda} \left( \int_{\sqrt{y_*}\lambda\cdot \omega}^\infty e^{-2\pi t^2} dt \right) e^{-\I \pi \tau_*\lambda^2 - \I
\textbf{w}\cdot \lambda} (-1)^{(\lambda-\lambda_0)\cdot w_2} & +  \\
+
2\pi \sum_{n=1}^\infty \frac{\left( \frac{\I \textbf{w}\cdot \omega}{2\sqrt{y_*}}\right)^n}{n!}
\sum_{\lambda} H_{n-1}(2\pi \sqrt{y_*} \lambda\cdot \omega) &
e^{-\I \pi \bar \tau_* \lambda_+^2 - \I \pi \tau_* \lambda_-^2  - \I
\textbf{w}\cdot \lambda} (-1)^{(\lambda-\lambda_0)\cdot w_2} \Biggr\} \\
\end{split}
\ee
where $\textbf{w}= \kappa_2 u^{1/6} S$, the constant $\kappa_2$ is given in equation \eqref{eq:kappa2-def},
and $H_n$ are standard Hermite polynomials.

In particular, if $\sigma = - 7$ so $\fB=0$ then we have a nonzero constant:
\be
G_\infty^{\omega} = - 2\pi \I \sum_{\lambda} \left( \int_{\sqrt{y_*}\lambda\cdot \omega}^\infty e^{-2\pi t^2} dt \right) e^{-\I \pi \tau_*\lambda^2 - \I
\textbf{w}\cdot \lambda} (-1)^{(\lambda-\lambda_0)\cdot w_2}
\ee
and if $\sigma = - 8$ so $\fB = 1$ then we the have a linear function of $S$:
\be
\begin{split}
G_\infty^{\omega} = - 2\pi \kappa_2  \Biggl\{ &
\sum_{\lambda} \left( \int_{\sqrt{y_*}\lambda\cdot \omega}^\infty e^{-2\pi r^2} dr \right)( S \cdot \lambda) e^{-\I \pi \tau_*\lambda^2 - \I
\textbf{w}\cdot \lambda} (-1)^{(\lambda-\lambda_0)\cdot w_2}\\
& +  \frac{ \pi }{\sqrt{y_*}} S\cdot \omega
\sum_{\lambda}   e^{-\I \pi \bar \tau_* \lambda_+^2 - \I \pi \tau_*\lambda_-^2 } (-1)^{(\lambda-\lambda_0)\cdot w_2}\Biggr\} \\
\end{split}
\ee
and so on. Clearly, these expressions depend continuously on the metric and do not vanish as $\omega$ approaches any boundary
of the light cone.

\section{The $U(1)_R$ Charge Anomaly}\label{sec:U1R-Anomaly}

There is a simple conceptual reason for the selection rule $U=0$ we have found:
It is the selection rule enforced by the the $U(1)_R$ symmetry of a superconformal theory.
As mentioned in the introduction, this is not surprising given the work of
\cite{Shapere:2008zf}.

It was pointed out that such $U(1)_R$ selection rules would apply to twisted superconformal correlators in
\cite{Marino:1998bm} although the background charge for the
AD3 theory deduced from the measure of the $SU(3)$ Coulomb branch was incorrectly stated in that paper to be $-\chi/10$.
The correct determination from the measure, expressed in terms of the conformal anomalies
$a$ and $c$, was given in \cite{Shapere:2008zf}. We briefly  recall the derivation here.

When an $N=2$ theory is coupled to external fields the
 anomaly for the $U(1)_R$  current can be deduced via the descent formalism from an index density in six dimensions.
 We introduce a $U(1)_R$ symmetry line bundle  $\CR$ with connection. Let $\CF_1$ be the fieldstrength of that connection.
Similarly we introduce a  principal $SU(2)_R$ symmetry bundle $P_R$. Let  $E$ denote the associated
bundle in the spinor representation.
\footnote{If $w_2(P_R)$ is nonzero we can make appropriate modifications by working in the adjoint representation.
But this normalization is the most convenient.}
It has a connection with fieldstrength $\CF_2$. In a Lagrangian theory we can write the relevant index density as:
\be
\begin{split}
I_6 & = \left[(\Tr e^{\frac{\CF_1\otimes 1 + 1\otimes \CF_2}{2\pi}} )  \hat A \right]_6 ~  . \\
\end{split}
\ee
where the trace is taken over the fermionic fields in the $N=2$ field multiplets with $\CF_1$ and $\CF_2$
in the corresponding representation. Expanding this out we get:

\be
I_6 = \Tr (T_{U(1)}^3 ) \frac{c_1(\CR)^3}{3!}  + \Tr( T_{U(1)} T_{SU(2)}^2) c_1(\CR) \ch_2(E) - \Tr(T_{U(1)}) c_1(\CR) \frac{p_1}{24}
\ee
where $T_{U(1)}$ is the generator of $U(1)_R$ symmetry and $T_{SU(2)}$ is any generator of the $SU(2)_R$ symmetry.

Now we use the relation between the $U(1)_R$
symmetry anomaly and the $a$ and $c$ coefficients of the
stress-tensor correlators, as derived in \cite{Anselmi:1997am,Anselmi:1997ys,Kuzenko:1999pi}.
These results are based on the structure of  superconformal
multiplets. (See \cite{Intriligator:2003jj} for a useful discussion.) The result is that
\be
\Tr (T_{U(1)})^3 = \Tr T_{U(1)} = 48 (a-c)
\ee
\be
\Tr \left( T_{U(1)}T^a_{SU(2)} T^b_{SU(2)} \right) = \delta_{ab} (4a-2c)
\ee
Substitution into the anomaly polynomial then expresses it in terms of the conformal anomalies $a,c$:
\be\label{eq:I6-ac}
I_6 = 2(a-c) \left( 4 c_1(\CR)^3 - c_1(\CR) p_1 \right) + 2(2a-c) c_1(\CR) \ch_2(E)
\ee
and the corresponding background charge computed via the descent formalism is:
\be
\Delta T_{U(1)} =  (a-c) \int_X \left(12 c_1(\CR)^2 -  p_1 \right) + 2(2a-c)  \int_X \ch_2(E) ~ .
\ee
Now, all three quantities $a,c$ and $I_6$  make sense in all $N=2$ theories, and in particular
in non-Lagrangian theories.  It is therefore natural to postulate that the expression for the
anomaly polynomial \eqref{eq:I6-ac} holds universally for all $N=2$, $d=4$ theories. We will
adopt this hypothesis. The computation in this paper can be viewed as a nontrivial check that
the hypothesis is correct.

Now in a twisted $N=2$ theory we have an isomorphism $E \cong S^+$, but
\be
\int_X \ch_2(S^\pm) = \frac{3\sigma \pm 2\chi}{2}
\ee
 and on any oriented four-manifold
$\int_X p_1 = 3 \sigma$. Putting these facts together we recover the result
of   \cite{Shapere:2008zf} that in a topologically twisted theory:
\footnote{Our normalization of the $U(1)_R$ charge differs by a factor of $2$ from that
of \cite{Shapere:2008zf}.}
\be
\Delta T_{U(1)} =     (2a-c) \chi + \frac{3}{2} c \sigma
\ee
Plugging in the values \cite{Aharony:2007dj,Shapere:2008zf,Xie:2013jc}   $ a= 43/120$ and $c=11/30$
leads to the specific result:
\be\label{eq:AD3-BackCharge}
\Delta T_{U(1)} =  \frac{7\chi + 11 \sigma}{20} ~ .
\ee
The sum of this value with the $R$-charges of the observables must vanish.
The $U(1)_R$ charge of the canonical $0$-observable is $6/5$ and hence that of the $2$-observable
$\CO(S)$ is $1/5$.
\footnote{To see this note that $\lambda$ and $Z$ have $U(1)_R$ charge $+1$, the supersymmetry operator
$K$ has charge $-\half$.}
Therefore, dividing the selection rule $U=0$, as found in our computations above,
by $5$ gives the expected $U(1)_R$ symmetry
selection rule:
\be\label{eq:RSelect-2}
\frac{6}{5} \ell+ \frac{1}{5} r = \frac{\chi_h - c_1^2}{5} = - \frac{1}{20} (7\chi + 11 \sigma)
\ee
in perfect harmony with \eqref{eq:AD3-BackCharge}.

\textbf{Remark}: When $b_1$ is nonzero we can also introduce $1$- and $3$-observables $\CO(\gamma)=\int_{\gamma} Ku $ and $\CO(\Sigma)=\int_{\Sigma} K^3 u $,  for $\gamma\in H_1(X;\IZ)$ and $\Sigma \in H_3(X;\IZ)$, respectively. The
selection rule now becomes
\be\label{eq:RSelect-2}
\frac{12}{10} n_0 + \frac{7}{10} n_1 + \frac{2}{10} n_2 - \frac{3}{10} n_3
=  \frac{\chi_h - c_1^2}{5} = - \frac{1}{20} (7\chi + 11 \sigma)
\ee
where $n_k$ is the number of insertions of the $k$-observable.
The notable feature here is that the relative minus sign in the sum on the
left-hand side allows the possibility of infinitely many nontrivial correlation functions.

\appendix

\section{The $u$-plane Integrand And Total Derivatives}\label{app:TotalDerivative}

In this appendix we  will show that if we consider the difference of
two $u$-plane measures at different period points $\omega$ and $\omega_0$
then the measure can naturally be written as a total derivative of a well-defined
one-form on the $u$-plane. Our approach here was strongly influenced
by the recent paper of Korpas and Manschot \cite{Korpas:2017qdo}.
The wall-crossing formula \eqref{eq:uPdiff} below is equivalent to that derived in
\cite{Moore:1997pc}.

Up to an overall constant the measure on the $u$-plane can be written as
\be\label{eq:mu-Coulomb}
d\mu_{\rm Coulomb}^{\omega} = du d \bar u \hat \CH  \hat\Psi
\ee
where (using $\chi + \sigma = 4$ for $b_2^+=1$)
\be
\hat \CH = \left(\frac{du}{da} \right)^{1-\sigma/2} \Delta^{\sigma/8} e^{2 p u + S^2 T}
\ee
is purely holomorphic, $\hat \Psi = \frac{d \bar a}{d\bar u} \Psi$, and
the period point $\omega$ is explicitly written in the notation in equation \eqref{eq:mu-Coulomb} because the
dependence of the measure on $\omega$ will be important in what follows.  We can
rewrite $\hat \Psi$ in a useful way as follows. Define
\be
\rho_\lambda^\omega:= \sqrt{y} \lambda_+ -  \frac{\I}{4\pi \sqrt{y} } S_+ \frac{du}{da}
\ee
so that
\be\label{eq:CN-deriv}
\CN_\lambda = - 4\I \left(  \frac{d \rho_\lambda^\omega}{d\bar a} e^{-2\pi (\rho_\lambda^{\omega})^2} \right) e^{-\I \pi \tau \lambda^2 - \I S \cdot \lambda
\frac{du}{da} }
\ee
Now define the entire function of $r$ and $b$:
\be\label{eq:Inc-Error}
\CE(r;b) := \int_b^r e^{-2 \pi t^2} dt
\ee
We will also denote $\CE(r):= \CE(r;0)$. It follows that  we can write
\be\label{eq:hatPsi-td1}
\hat\Psi = -4\I \sum_{\lambda} \left( \frac{d}{d\bar u} \CE(\rho^\omega_{\lambda}; b_\lambda) \right)
e^{-\I \pi \tau \lambda^2 - \I S \cdot \lambda
\frac{du}{da} }(-1)^{(\lambda-\lambda_{0,\infty})\cdot \xi_\infty}
\ee
Here the lower bounds $b_\lambda$ in the contour integral \eqref{eq:Inc-Error} are fairly arbitrary. They can
depend on $\lambda$ and $u$, but not on $\bar u$.

Given the expression \eqref{eq:hatPsi-td1} and the fact that $\hat \CH$ is holomorphic one is strongly
tempted to write the  $u$-plane integrand as a total derivative
\be\label{eq:TotalDeriv}
d\mu_{\rm Coulomb}^{\omega} = d \Omega
\ee
where $\Omega$ is a $(1,0)$ form:
\be\label{eq:Total-1}
\Omega = - du \hat \CH \widetilde{\Theta}
\ee

In this expression we introduced an indefinite theta function:
\be\label{eq:FirstTildeTheta}
\widetilde{\Theta} =
\widetilde {\Theta}(\xi, \lambda_0; \tau, \textbf{z}; \{ b_\lambda \}  ) :=   \sum_{\lambda\in \lambda_0 + \Gamma }
\CE(\rho^\omega_{\lambda}(\textbf{z}); b_\lambda) e^{-\I \pi \tau \lambda^2 - 2\pi \I \textbf{z}\cdot \lambda  }e^{ 2 \pi \I (\lambda-\lambda_0)\cdot \xi}
\ee
where
\be
\rho^\omega_{\lambda}(\textbf{z}):=  \sqrt{y} (\lambda - \frac{\I\textbf{z}}{2y}  ) \cdot \omega
\ee
and the lower bounds $\{ b_\lambda \}$  should be chosen so that
 the summation is absolutely convergent. Note that the factor
\be
e^{-\I \pi \tau \lambda^2} = e^{\pi y \lambda^2} e^{-\I \pi x \lambda^2}
\ee
can potentially lead to an exponential divergence from an infinite sum of vectors $\lambda$ with $\lambda^2 \to + \infty$,
so the constants $\{ b_\lambda \}$ must be chosen so that the error function decays fast enough to overwhelm this
potential divergence.

The one-form in \eqref{eq:Total-1} uses the function $\tilde\Theta$ with   $\Gamma= \bar H^2(X)=H^2(X)/{\rm Tors}$ and
\be\label{eq:z-def}
\textbf{z} = \frac{1}{2\pi} S \frac{du}{da} ~ .
\ee

The problem with the expression \eqref{eq:FirstTildeTheta} is that there is a conflict between
absolute convergence and single-valuedness of $\Omega$ on the $u$-plane. There are choices of
$\{ b_\lambda \}$, e.g. $b_\lambda =0$ for which a \underline{formal} application of the
Poisson summation formula would prove that $\Omega$ is single-valued, but such a choice
leads to divergences since if we take $b_\lambda=0$ then
\be
\CE(\rho^\omega_\lambda) \rightarrow \frac{1}{\sqrt{8}} {\rm sign}(\lambda_+)
\ee
for $\lambda_+ \to \infty$. In fact, if there were a one form $\Omega = - du \hat\CH F$
such that \eqref{eq:TotalDeriv} holds and such that   $\Omega$ is single-valued on the $u$-plane then we would have
\be
0  = \oint d\bar u \frac{d}{d\bar u} (\hat \CH F ) = \oint d \bar u \hat \CH \hat \Psi
\ee
around any closed path in the $u$-plane. In particular this includes paths along which
the monodromy of the local system of electro-magnetic charges is nontrivial. It is
easy to check that in general the relevant periods are nonzero.

The situation is quite different if we consider instead the \underline{difference} of
$u$-plane measures for two different metrics with period points $\omega$ and $\omega_0$.
Then we can indeed write
\be\label{eq:DiffTotDeriv}
d\mu_{\rm Coulomb}^{\omega} - d\mu_{\rm Coulomb}^{\omega_0}  = d \Omega^{\omega,\omega_0}
\ee
where $\Omega^{\omega,\omega_0}$ is a $(1,0)$ form:
\be\label{eq:Total-2}
\Omega^{\omega,\omega_0} = - du \hat \CH \widetilde{\Theta}^{\omega,\omega_0}
\ee

In this expression we make use of the general indefinite theta series
\footnote{
We remark that the   function \eqref{eq:SecondTildeTheta} is similar to, but different from
that discussed in \cite{Korpas:2017qdo,Zwegers,Vigneras}. The
difference is that in the error function we do not take the imaginary part of $\textbf{z}$.}
\be\label{eq:SecondTildeTheta}
\widetilde {\Theta}^{\omega, \omega_0}(\xi, \lambda_0; \tau, \textbf{z} ) :=   \sum_{\lambda\in \lambda_0 + \Gamma }
\CE(\rho^{\omega}_{\lambda}(\textbf{z}), \rho^{\omega_0}_\lambda(\textbf{z}) )
 e^{-\I \pi \tau \lambda^2 - 2\pi \I \textbf{z}\cdot \lambda  }e^{ 2 \pi \I (\lambda-\lambda_0)\cdot \xi}
\ee
It is both absolutely convergent and satsifies the modular transformation properties
\be\label{eq:tauplusone}
\widetilde {\Theta}^{\omega, \omega_0}(\xi, \lambda_0; \tau+1, \textbf{z} ) = e^{-\I \pi \lambda_0^2}
\widetilde {\Theta}^{\omega, \omega_0}(\xi-\half (w_2 + 2\lambda_0) , \lambda_0; \tau, \textbf{z} )
\ee
\be\label{eq:ModTmn}
\widetilde {\Theta}^{\omega, \omega_0}(\xi, \lambda_0; -\frac{1}{\tau}, \frac{ \textbf{z}}{\tau} )
= e^{ \I \pi/2}  (-\I \tau)^{d/2} e^{-2\pi\I \lambda_0 \cdot \xi}
e^{-\I\pi \textbf{z}^2/\tau}  \widetilde {\Theta}^{\omega, \omega_0}(\lambda_0, -\xi;\tau,  \textbf{z}  )
\ee
where $d$ is the rank of $\Gamma$.

Now, in \eqref{eq:Total-2} we use the function \eqref{eq:SecondTildeTheta} with $\textbf{z}$ given
in \eqref{eq:z-def}. It is straightforward to show  that $\Omega^{\omega,\omega_0}$ is single-valued on the $u$-plane.

It now follows that the difference of $u$-plane integrals can be written as:
\be\label{eq:uPdiff}
\begin{split}
Z_u^{\omega} - Z_u^{\omega_0} & = 4 \I e^{2\pi \I \lambda_{0,\infty}^2 } K_u
\int_{\rm u-plane} d \left(   du \hat\CH \widetilde{\Theta}^{\omega,\omega_0}(\xi,\lambda_0;\tau, \textbf{z}  \right) \\
& = 4 \I e^{2\pi \I \lambda_{0,\infty}^2 } K_u \lim_{\epsilon\to 0}
\left[ \oint_{\vert u \vert = 1/\epsilon } du \hat\CH \widetilde{\Theta}^{\omega,\omega_0}
- \sum_{s} \oint_{\vert u - u_s \vert = \epsilon } du  \hat \CH \widetilde{\Theta}^{\omega,\omega_0} \right] \\
\end{split}
\ee
The contour integrals are all oriented counterclockwise. This is a somewhat better way of phrasing the
wall-crossing formula presented in \cite{Moore:1997pc}.

We now use the contour integral representation \eqref{eq:uPdiff} to show that there is a
formal series in $p,S$ expressed as a \underline{contour integral}, and
denoted $G^{\omega}(p,S)$, or just $G^{\omega}$, such that
\be\label{eq:DeltaZ-DeltaG}
Z_u^\omega - Z_u^{\omega_0} = G^{\omega} - G^{\omega_0} ~ .
\ee
The point here is that $G^{\omega}$ only depends on a single period point, and yet it is expressed
as a contour integral.
Before deriving \eqref{eq:DeltaZ-DeltaG} let us draw from it some useful consequences.

First, \eqref{eq:DeltaZ-DeltaG} implies  that $Z_u^\omega = G^{\omega} + constant$ where
the ``constant'' does not depend on $\omega$ but can be a power series in $p$ and $S$.
As we will see in the derivation of \eqref{eq:DeltaZ-DeltaG}  the
 formula is only valid when $\omega$ and $\omega_0$ are
in the same component of the light cone in $H^2(X;\IR)$. On the
other hand,  $Z_u^\omega$ is defined for $\omega$ in either component
and moreover $Z_u^{-\omega} = - Z_u^{\omega}$. Therefore we can conclude that
\be
Z_u^\omega = G^{\omega} + C(p,S)  {\rm sign}(\omega^t)
\ee
where $C(p,S)$ is independent of $\omega$ and $\omega^t$ is the ``time component'' of $\omega$.

It is, unfortunately, difficult to
give useful explicit expressions for $C(p,S)$. However, there is one case
in which we can be more definitive. For $X$ of a suitable homotopy type
there are vanishing chambers for $Z_u^{\omega}$ in the sense explained in sections 5 and 6 of  \cite{Moore:1997pc}.
That is, for any monomial $p^\ell S^r$ in the power series there is a region
$\CV_{\ell, r}$ near the boundary of the light cone so that the contribution
of $Z_u^\omega$ to that monomial vanishes for $\omega \in \CV_{\ell, r}$. Moreover the
regions form an inverse system: There is an ordering so for $(\ell',r') > (\ell, r)$
$\CV_{\ell',r'} \subset \CV_{\ell,r}$.
Let $\CV$ be the inverse limit of these vanishing chambers so we can say
that $Z_u^{\omega_0} = 0 $ for $\omega_0 \in \CV$. This simply means that
the coefficient of any monomial $p^\ell S^r$ in  $Z_u^{\omega_0}$
vanishes for $\omega_0 \in \CV_{\ell', r'} $ for $(\ell',r') > (\ell, r)$. In this
sense it follows  from \eqref{eq:DeltaZ-DeltaG} that
\be\label{eq:Zuomega-contour}
Z_u^\omega  = G^{\omega} - G^{\omega_0}    \qquad\qquad  \omega_0 \in \CV ~ .
\ee
Therefore for such $X$ we can express $Z_u^\omega$ as a sum of contour integrals around
the singular points. The class of homotopy types for which this applies is rather broad.
It includes rational surfaces and blow-ups of surfaces for suitable choices of $\lambda_{0,\infty}$.
In particular it applies to such manifolds for the main example of this paper, where
$\lambda_{0,\infty} = \xi_\infty = \lambda_0 = \half \overline{w_2(X)}$.

It remains to prove  \eqref{eq:DeltaZ-DeltaG}. We begin with the contribution of a finite point $u_s$ and assume that
$\Im \tau_s \to \infty$ and  $da_s/du$ is a finite period as $u\to u_s$.
Then, provided the metric is generic so that there is no $\lambda$ with $\lambda_+=0$,
equation \eqref{eq:SecondTildeTheta} simplifies and we can replace the difference of
error functions by
\be
\frac{1}{\sqrt{8}}\left(
{\rm sign}(\lambda\cdot \omega) - {\rm sign}(\lambda\cdot \omega_0)\right)
\ee
But now we note that if $\omega$ and $\omega_0$ are in
the same component of the lightcone then  their time components $\omega^t$
and $\omega_0^t$ have the same sign and hence
\be
{\rm sign}(\lambda\cdot \omega) - {\rm sign}(\lambda\cdot \omega_0)=0
\ee
when $\lambda^2 \geq 0$. Therefore, in evaluating the residue integral  $\Delta Z_{u,s}^{\omega,\omega_0}$
around $u_s$ in \eqref{eq:uPdiff} we can
replace $\widetilde{\Theta}^{\omega, \omega_0}$ by $F_s^{\omega} - F_s^{\omega_0}$ where
we define:
\be\label{eq:FirstModifiedTheta}
F_s^{\omega}  :=  \frac{1}{\sqrt{8}} \sum_{ \lambda:  \lambda^2 < 0  }
{\rm sign}(\lambda\cdot \omega)  e^{-\I \pi \tau \lambda^2 - 2\pi \I \textbf{z}\cdot \lambda  }e^{ 2 \pi \I (\lambda-\lambda_0)\cdot \xi}
\ee
and $\textbf{z}$ is defined as in \eqref{eq:z-def}. Note carefully that because of the restriction $\lambda^2<0$   this sum
converges absolutely. Moreover it is a function purely of $\omega$ and not of $\omega_0$. Let $G_s^{\omega}$ be the corresponding
contour integral
\be
G^{\omega}_s := \oint_{u_s} du \hat \CH F_s^\omega  ~~ .
\ee

We would like to do something similar to write $\Delta Z_{u,\infty}^{\omega,\omega_0} = F_\infty^{\omega} - F_\infty^{\omega_0}$
but here we have two complications:

\begin{enumerate}

\item  For $SU(2)$ $N_f < 4$ we have $du/da \to \infty$ as $u\to \infty$.

\item For the conformal theories of interest we have $\tau \to \tau_*$ as $u\to \infty$ and $Im(\tau)$ does not go to infinity
so we cannot replace the error functions by differences of sign functions.

\end{enumerate}

To deal with these complications we note that the $u$-plane integral really only has meaning as a formal power series in $p$ and $S$.
Therefore, we should use the expansion  of the error function
\be\label{eq:ErrorFunExpan}
\CE(r+a) = \CE(r) - 2\pi e^{-2\pi r^2}  \sum_{n=1}^\infty \frac{(-2\pi a)^n}{n!} H_{n-1}(2\pi r)
\ee
where $H_n(x)$ are the standard Hermite polynomials. We  apply \eqref{eq:ErrorFunExpan}  with $r= \sqrt{y} \lambda\cdot \omega$ and
$ a= - \frac{\I}{4\pi \sqrt{y}} \frac{du}{da} S\cdot \omega$.   This gives:
\be\label{eq:ThetaExpError}
\begin{split}
\widetilde {\Theta}^{\omega, \omega_0}(\xi, \lambda_0; \tau, \textbf{z} ) &  :=   \sum_{\lambda\in \lambda_0 + \Gamma }
\left[ \CE(\sqrt{y}\lambda \cdot \omega ) -
\CE(\sqrt{y} \lambda \cdot \omega_0 ) \right] e^{-\I \pi \tau \lambda^2 - 2\pi \I \textbf{z}\cdot \lambda  }e^{ 2 \pi \I (\lambda-\lambda_0)\cdot \xi} \\
& +
\sum_{n=1}^\infty (\Theta^{\omega}_n - \Theta^{\omega_0}_n ) \\
\end{split}
\ee
where the $\Theta_n^{\omega}$ come from the $n^{th}$ term in the sum in \eqref{eq:ErrorFunExpan} and
are absolutely convergent sums on $\lambda$.
For a fixed monomial $p^\ell S^r$  only a finite number of such terms will contribute
so we do not need to worry about the convergence of the sum on $n$ in $\sum_n \Theta_n$.
Now, since we are considering a contour on a circle whose radius goes to infinity,
if $y \to \infty$ we can replace this expression  by $F_\infty^\omega - F_\infty^{\omega_0}$
where
\be
F_\infty^\omega:=  \sum_{\lambda\in \lambda_0 + \Gamma,~  \lambda^2 < 0  }
{\rm sign}(\lambda\cdot \omega)  e^{-\I \pi \tau \lambda^2 - 2\pi \I \textbf{z}\cdot \lambda  }e^{ 2 \pi \I (\lambda-\lambda_0)\cdot \xi}
+ \sum_{n=1}^\infty \Theta_n
\ee
is a well-defined function of a single period point $\omega$.

In the conformal case where $y\to y_*$ has a finite limit as $u\to \infty$ we write
\be
 \CE(\sqrt{y}\lambda \cdot \omega ) -
\CE(\sqrt{y} \lambda \cdot \omega_0 )  =  \CE(\sqrt{y}\lambda \cdot \omega;\infty ) -
\CE(\sqrt{y} \lambda \cdot \omega_0;\infty )
\ee
Now we can separate terms and obtain a well-defined function $F_\infty^{\omega}$.

Finally, let $G_\infty^{\omega}$ denote the contour integral of $du \hat \CH F_\infty^{\omega}$ around the circle
at infinity and let
\be
G^{\omega} : = G_\infty^{\omega} + \sum_s G_s^{\omega}.
\ee
This completes the proof of \eqref{eq:DeltaZ-DeltaG}.

\section{Detailed Derivation Of The Relation Of $SU(2)$, $N_f=1$ And $AD3$ $u$-plane Integrals}\label{app:Zutilde}

In this appendix we prove the crucial claims made between equations \eqref{eq:CommLim} and
\eqref{eq:MainConjecture} for the \underline{difference} of $u$-plane integrals for different
period points $\omega,\omega_0$. Using the results of Appendix \ref{app:TotalDerivative}
we see that if we take the difference of the quantity in
equation \eqref{eq:LimInt-ne-IntLim} for two period points then it can be written as a sum of contour integrals.

  We consider a small disk $B(u_*;\epsilon)$ of
radius $\epsilon$ around the critical point $u_*$. Let $\gamma_\epsilon$ be the counterclockwise oriented
boundary. Set $\Lambda_1 = 1$ so that $u_* = 1$ and define the deviation from the critical mass by $m= \frac{3}{2} + \mu$.
Then we cut out disks of radius $\delta$, with $\delta \ll \epsilon$  around the colliding points $u_\pm$ in the discriminant locus
and let $\gamma_\pm$ be the ccw oriented boundaries of these disks. We are going to prove that
\be
P_1(p,S):= \left[ e^{-2p \left( u_* + \frac{2}{3}\mu \right) - T_* S^2} \left(
\oint_{\gamma_\epsilon} \Omega -  \oint_{\gamma_+} \Omega - \oint_{\gamma_-} \Omega \right) \right]_{\mu^0}
\ee
is a polynomial in $p$ and $S$. Here it is understood that we take $\delta \to 0$ then $\epsilon\to 0$.
As mentioned above the quantity in square brackets might have divergent terms for $\mu \to 0$. It has a Laurent
expansion in $\mu^{1/4}$ around $\mu=0$. The singular terms will  cancel against terms coming from the
Seiberg-Witten contribution to the partition function. In any case, our main focus here is on the
constant term, i.e. the coefficient of $\mu^0$.

Moreover, we will compare the polynomial $P_1(p,S)$  to the $u$-plane contribution for the AD3 theory
\be
P_{AD}(p,S):= \left[  \left(
\oint_{\gamma_\infty} \Omega -  \oint_{\gamma_+^{AD}} \Omega - \oint_{\gamma_-^{AD} } \Omega \right) \right]_{\Lambda_{AD}^0}
\ee
where now $\gamma_\pm^{AD}$ are small contours of radius $\epsilon$ around the two points in the AD3 discriminant
locus $u_\pm = \pm 2 \Lambda_{AD}^3$. We will show that $P_{AD}(p,S)$ is also a polynomial in $p$ and $S$.
Furthermore,  if we define a grading of the polynomial $P_1$ by ``$R$ charge'' with
 $R[p]= 6$ and $R[S]=1$ then we will show that the highest degree is given by $6\ell + r = \fB = -\frac{1}{4}(7\chi + 11 \sigma)$.
 Finally,   defining $P_1^{\rm top}(p,S)$ be be the sum of terms with maximal
 $R$-charge we will show that
\be\label{eq:Rescalings1}
P_1^{\rm top}(p,S) = N P_{AD}(n_0 p, n_2 S)
\ee
for suitable constants $N,n_0,n_2$.

In the proof it is useful to note that for $b_2^+=1$ we have $\fB = - 7-\sigma $ and
 $1-\chi/2 = \sigma/2 - 1 $ and we recall that, up to an overall normalization we have
 \eqref{eq:DiffTotDeriv} with
\be
\Omega = du \left( \frac{du}{da} \right)^{1-\sigma/2} \Delta^{\sigma/8} e^{2pu + S^2 T}
\widetilde {\Theta}^{\omega, \omega_0}(\lambda_0, \lambda_0; \tau, \textbf{z} )
\ee
where $\lambda_0 = \half \overline{w_2(X)}$.
It will be crucial to compare expressions for $du/da$ and $u$ in the relevant expansions
in the $N_f=1$ and $AD3$ contour integrals.

We begin with the expression in the $N_f=1$ theory
\be\label{eq:Nf-gammaepsilon}
\left[ e^{-2p \left( u_* + \frac{2}{3}\mu \right) - T_* S^2}
\oint_{\gamma_\epsilon} \Omega  \right]_{\mu^0}
\ee
 Here we can set $\mu=0$ in the expressions for $\Omega$
so that the two points $u_\pm$ collide at $u=u_*$. In evaluating this integral
we expand the integrand in powers of $(u-u_*)$ and perform the contour integral.
When $\mu=0$ we find that $\tau(u)$ approaches $\tau_* = e^{\I \pi/3}$ as
$u\to u_*$ and indeed
\be\label{eq:ttstar}
\tau = \tau_* + PS((u-u_*)^{1/3} )
\ee
where $PS(x)$ means a power series in positive powers of $x$ that vanishes at $x=0$.
Similarly:
\be\label{eq:duda-Nf1-epsilon}
\left(\frac{du}{da} \right) =  \kappa_1 (u-u_*)^{1/6} \left( 1+ PS((u-u_*)^{1/3} ) \right)
\ee
with
\be
\kappa_1 = \left( - \frac{1}{4} \left( \frac{3}{\rho} \right)^2\left( - \frac{4}{9} \right)^{1/3}
(E_6(\tau_*))^{-1/3} \right)^{1/2}
\ee
Similarly,
\be
du \left( \frac{du}{da} \right)^{1-\sigma/2} \Delta^{\sigma/8} = \CN^1_{\infty} \frac{d(u-u_*)}{(u-u_*)} (u-u_*)^{-\fB/6} (1 + PS((u-u_*)^{1/3}) )
\ee
with
\be
\CN^1_{\infty} = \kappa_1^{1-\sigma/2} (u_*-u_0)^{\sigma/8}
\ee
and finally, $T_* = u_*/3$ and
\be
T-T_* = - \frac{\kappa_1^2}{24} E_2(\tau_*) (u-u_*)^{1/3} \left( 1+ PS( (u-u_*)^{1/3}) \right)
\ee

The integral over the phase of $u-u_*$ will kill all terms in the power series except those proportional to
\be
\frac{d(u-u_*)}{(u-u_*)}( \vert (u-u_*)^{1/3} \vert^2)^n
\ee
for some integer $n$, and in our expressions $n$ is always nonnegative.
 However, since we also take the $\epsilon\to 0$ limit, only the terms with $n=0$ will
contribute. We thus concentrate on the Laurent expansion in  $(u-u_*)^{1/3}$ working to zeroth order in the
power series expansion in $(\bar u - \bar u_*)^{1/3}$.

Now, since $(\tau-\tau_*)$ and  $du/da$ are expansions in positive powers of $(u-u_*)^{1/3}$ the resulting contour integral
is a polynomial in $p$ and $S$. Moreover, $S$ always multiplies $du/da$, so by
\eqref{eq:duda-Nf1-epsilon} if we assign charge $+1$ to $S$  and  $+6$ to $p$ then the leading powers of
$(u-u_*)^{1/3}$ are governed by the natural grading $6\ell + r$.
The higher order terms in the expansions in $(u-u_*)^{1/3}$ above  will contribute
to lower degree terms in the polynomial $P_1$. So the contribution to $P_1^{\rm top}$ only comes
from the leading order terms in the above expansions giving the contribution to the polynomial:
\be
P_{1,\infty}^{\rm top}(p,S)=
\CN^1_\infty \oint \frac{d(u-u_*)}{(u-u_*)} (u-u_*)^{-\fB/6} e^{2p(u-u_*)}  F_\infty(\kappa_1 (u-u_*)^{1/6} S)
\ee
where
\be\label{eq:F-infty-def}
F_\infty(\textbf{w}) = e^{-\frac{\textbf{w}^2}{24} E_2(\tau_*)}
\sum_{\lambda\in \Gamma+\lambda_0} (-1)^{w_2\cdot(\lambda-\lambda_0)}
 \CE\left(\rho^{\omega}_\lambda(\textbf{w}); \rho^{\omega_0}_\lambda(\textbf{w}) \right)
e^{-\I \pi \tau_* \lambda^2 - \I \textbf{w} \cdot \lambda }
\ee
and here
\be
\rho_\lambda^{\omega}(\textbf{w}):
= \sqrt{y_* } \lambda_+ -  \frac{\I}{4\pi \sqrt{y_*} }\textbf{w}_+ ~~ .
\ee

Let us compare the above contribution to $P_1^{\rm top}$  with the corresponding expression in the
$AD3$ theory
\be
\left[\oint_{\gamma_\infty} \Omega^{AD}\right]_{\Lambda_{AD}^0}
\ee

Since we are after the constant term we consider the  AD3 family with $\Lambda_{AD} \to 0$.
Equation \eqref{eq:E6E4g3g2} can be written as:
\be\label{eq:E6E4g3g2-2}
\frac{ (E_4(\tau))^3}{(E_6(\tau))^2}= 4 \left( \frac{\Lambda^3}{u} \right)^2
\ee
and \eqref{eq:dadu} can be written as
\be\label{eq:E6Edadu-2}
\left( \frac{du}{da} \right)^2 = - \frac{1}{6} \left( \frac{3}{\rho} \right)^2 \frac{E_4(\tau)}{E_6(\tau)} \frac{u}{\Lambda^2}
\ee
Now, as $u \to \infty$,
\be
\tau-\tau_* = 2^{2/3} \frac{(E_6(\tau_*))^{2/3}}{E_4'(\tau_*)}\left( \frac{\Lambda^3}{u} \right)^{2/3}\left( 1+ PS( \left( \frac{\Lambda^3}{u} \right)^{2/3} )\right)
\ee
and
\be
\left(\frac{du}{da} \right)= \kappa_2 u^{1/6}
\left( 1+ PS( \left( \frac{\Lambda^3}{u} \right)^{2/3} )\right)
\ee
\be\label{eq:kappa2-def}
\kappa_2 = \left(  - \frac{1}{12} \left( \frac{3}{\rho} \right)^2
\frac{2^{2/3}}{(E_6(\tau_*))^{1/3} } \right)^{1/2}
\ee
Similarly,
\be
du \left( \frac{du}{da} \right)^{1-\sigma/2} \Delta^{\sigma/8} = \CN^{AD}_\infty \frac{d u}{u} u^{-\fB/6} (1 + PS( \left( \frac{\Lambda^3}{u} \right)^{2/3} )  )
\ee
with
\be
\CN^{AD}_\infty= \kappa_2^{1-\sigma/2}
\ee

Once again, since we are taking the contour to infinity, we can focus on the holomorphic expansion
in $u^{1/6}$.
All the higher order terms in the power series have positive powers of $\Lambda_{AD}$ and hence,
again, we need only consider the leading order terms to get the contribution at $\Lambda_{AD}^0$.
We have
\be
P_{AD,\infty}(p,S) = \CN^{AD}_\infty \oint_{\infty} \frac{d u}{u} u^{-\fB/6} e^{2pu} F_\infty(\kappa_2 u^{1/6} S)
\ee
with the same function $F_{\infty}$ defined in \eqref{eq:F-infty-def}.

Comparing the two expressions we will find an equality of the
kind \eqref{eq:Rescalings1}, for this contribution to the polynomial,
provided
\be
\CN^1_\infty (2p)^\ell (\kappa_1 S)^r = N \CN^{AD}_{\infty} (2 n_0 p)^{\ell} (n_2\kappa_2 S)^r
\ee
for $r + 6\ell = \fB$. We solve for $r$ in terms of $\ell$ and $\fB$ and then since different
powers of $\ell$ appear in the polynomial we must have
\be\label{eq:CompareCond-1}
\CN^1_\infty \kappa_1^{\fB}  = N \CN^{AD}_\infty (n_2 \kappa_2)^{\fB}
\ee
\be\label{eq:CompareCond-2}
\left( \frac{\kappa_2}{\kappa_1} \right)^6 = \frac{n_0}{n_2^6}
\ee

Now we consider an analogous computation for the contributions from $\gamma_\pm$.
First we consider
\be
\left[ e^{-2p \left( u_* + \frac{2}{3}\mu \right) - T_* S^2} \left(   \oint_{\gamma_\pm} \Omega \right) \right]_{\mu^0}
\ee
in the $N_f=1$ theory. Here we will be writing the integrand as a power series in the local duality frame variable $q_\pm$.

For small $\mu$ the two points in the discriminant locus have an expansion
\be
\begin{split}
u_+ & =1+ \frac{2}{3} \mu + \left( \frac{2}{3}   \right)^{5/2}\mu^{3/2} + \cdots \\
u_- & =1+ \frac{2}{3} \mu - \left( \frac{2}{3} \right)^{5/2}\mu^{3/2} + \cdots \\
\end{split}
\ee
A subtle point is that if we take the limit as $u\to u_*$ with $\mu$ held fixed then
the expansions for $u$ and $du/da$ involve   an infinite series of increasingly divergent terms in $\mu$.
The correct scaling
limit
\footnote{This is a consequence of the linear combinations we found in equation \eqref{eq:Nf1-to-AD3} above.}
is to define
\be
u= u_\pm + \mu^{3/2} v
\ee
and take the limit $\mu \to 0$ holding $v$ fixed. With this understood we have
\be
e^{2p u } = e^{2p (u_* + \frac{2}{3}\mu)} e^{\pm 2p \mu^{3/2} (2/3)^{5/2} E_6/E_4^{3/2}(1+\CO(\mu^{1/2})) }
\ee
where the Eisenstein series are expansions in $q_\pm$ in the standard way.
Next we can write
\be
 \frac{du}{da}  = \kappa_3   E_4^{-1/4} \mu^{1/4} \left( 1+ PS(\mu^{1/2}) \right)
\ee
\be
\kappa_3 =   \left(\frac{\zeta_s}{2} \left( \frac{3}{\rho} \right)^2 \sqrt{\frac{2}{27}} \right)^{1/2}
\ee
and similarly
\be
du \left( \frac{du}{da} \right)^{1-\sigma/2} \Delta^{\sigma/8} = \CN_\pm^1 \mu^{-\fB/4} \frac{d q_\pm }{q_\pm }
H(q_\pm) \left( 1 + PS(\mu^{1/2}) \right)
\ee
where the power series in $\mu^{1/2}$ has coefficients which are themselves power series in $q_\pm$.
Here
\be
H(q) := \left( q \frac{d}{dq} (\frac{E_6}{E_4^{3/2}}) \right) E_4^{-(\sigma+1)/4}\left(E_6^2-E_4^3\right)^{\sigma/8}
\ee
\be
\CN_\pm^1 = \pm \left( \frac{2}{3} \right)^{5/2(1+\sigma/4) } \kappa_3^{1-\sigma/2} (u_*-u_0)^{\sigma/8}
\ee

Now, the expansion in $p^\ell S^r$ comes with a power $\mu^{(r + 6 \ell)/4}$ so the $\mu^0$ term
satisfies the selection rule and the higher powers in the $\mu$ expansion contribute lower order terms.
Thus, the contribution to the polynomial from these two singularities is the sum over $+$ and $-$ of
\be
P_{1,\pm}^{\rm top}(p,S) = \eta \CN_\pm^1 \left[
\mu^{-\fB/4}\oint \frac{d q_\pm }{q_\pm } H(q_\pm)
e^{\pm 2p \mu^{3/2} (2/3)^{5/2} E_6/E_4^{3/2}  }
F_\pm(\kappa_3 \mu^{1/4} E_4^{-1/4} S ) \right]_{\mu^0}
\ee
where
\be\label{eq:F-pm-def}
F_\pm(\textbf{w})
 =\frac{1}{\sqrt{8}} e^{-\frac{\textbf{w}^2}{24} E_2(\tau)}
\sum_{\lambda\in \Gamma+\lambda_0} (-1)^{w_2\cdot(\lambda-\lambda_0)}
 \left( {\rm sign}(\lambda\cdot \omega ) - {\rm sign}(\lambda\cdot \omega_0)   \right)
e^{-\I \pi \tau \lambda^2 - \I \textbf{w} \cdot \lambda }
\ee

Finally we come to the contributions
\be
\left[\oint_{\gamma_\pm^{AD}} \Omega^{AD}\right]_{\Lambda_{AD}^0}
\ee
in the AD3 theory.

In the  AD3 theory we have the exact formulae for the expansions in $q_\pm$ near $u_\pm$:
\be
u = \pm 2 \Lambda_{AD}^3 \frac{E_6}{E_4^{3/2}}
\ee
\be
 \frac{du}{da} = \kappa_4  E_4^{-1/4} \Lambda_{AD}^{1/2}
\ee
\be
\kappa_4 = \left( - \zeta_s \frac{1}{6} \left( \frac{3}{\rho} \right)^2 \right)^{1/2}
\ee
and we compute:
\be
du \left( \frac{du}{da} \right)^{1-\sigma/2} \Delta^{\sigma/8} = \CN_\pm^{AD} \Lambda_{AD}^{-\fB/2} \frac{d q_\pm }{q_\pm }
H(q_\pm)
\ee
\be
 \CN_\pm^{AD} = \pm 2^{1+\sigma/4} \kappa_4^{1-\sigma/2}
\ee
So these terms contribute to the polynomial
\be
P_{AD,\pm} = \eta \CN_\pm^{AD} \left[ \Lambda_{AD}^{-\fB/2} \oint \frac{d q_\pm }{q_\pm }
H(q_\pm)e^{\pm 4 p  \Lambda_{AD}^3 \frac{E_6}{E_4^{3/2}}  }
F_\pm(\kappa_4 \Lambda_{AD}^{1/2} E_4^{-1/4}  S) \right]_{\Lambda_{AD}^0}
\ee

Now to match these using the rescalings \eqref{eq:Rescalings1} we have the conditions
\be
\CN^1_+ \left( 2 p \left(\frac{2}{3} \right)^{5/2} \right)^\ell (\kappa_3 S)^r =
N \CN^{AD}_+ (4n_0 p)^\ell (\kappa_4 n_2 S)^r
\ee
when $6\ell + r = \fB$. In a way similar to \eqref{eq:CompareCond-1} and \eqref{eq:CompareCond-2}
we obtain:
\be\label{eq:CompareCond-3}
\CN^1_+ \kappa_3^{\fB}  = N \CN^{AD}_+ (n_2 \kappa_4)^{\fB}
\ee
\be\label{eq:CompareCond-4}
\left( \frac{\kappa_4}{\kappa_3} \right)^6 = 2 \left( \frac{3}{2} \right)^{5/2} \frac{n_0}{n_2^6}
\ee

We now ask if there are constants $N, n_0,n_2$ that allow us to solve the four conditions
\eqref{eq:CompareCond-1}\eqref{eq:CompareCond-2}\eqref{eq:CompareCond-3}\eqref{eq:CompareCond-4}. The conditions
are not all independent, and in fact, there are such constants iff we have
\be
\left( \frac{\kappa_1 \kappa_4 }{\kappa_2 \kappa_3} \right)^6= 2^{-3/2} 3^{5/2}
\ee
\be
\frac{\CN^1_\infty}{\CN^1_+}  \left( \frac{\kappa_1}{\kappa_3} \right)^{\fB}= \frac{\CN^{AD}_\infty}{\CN^{AD}_+} \left( \frac{\kappa_2}{\kappa_4} \right)^{\fB}
\ee

Plugging in the above values we can confirm that these conditions are indeed satisfied.

\end{document}